\documentclass[twocolumn,showpacs,preprintnumbers,amsmath,amssymb,floatfix,nofootinbib]{revtex4-1}
\usepackage{color}   
\usepackage{graphicx}
\usepackage{dcolumn}
\usepackage{bm}
\usepackage{slashed}
\usepackage{float}
\usepackage{graphicx,changepage}
\usepackage{ulem}    

\begin{document}

\def\bb    #1{\hbox{\boldmath${#1}$}}


\twocolumngrid

\title{On the Clustering Properties of Produced Particles  in
  High-Energy $pp$ Collisions}

\author{Cheuk-Yin Wong$^1$, Hanpu Jiang$^2$, Nanxi Yao$^2$,
  Liwen Wen$^2$, Gang Wang$^2$, Huan Zhong Huang$^{2,3}$}
\affiliation
    {$^1$Physics Division, Oak Ridge National Laboratory,
      Oak  Ridge, TN 37831, USA \\
$^2$Department of Physics and Astronomy,
  University of California, Los Angeles, California 90095, USA\\
  $^3$Key Laboratory of Nuclear Physics and Ion-beam Application
  (MOE) and Institute of Modern Physics, \\Fudan University, Shanghai
  200433, China }


\begin{abstract}

Minijets provide useful information on parton interactions in the low
transverse-momentum\break  (low-$p_T$) region.  Because minijets produce
clusters, we study the clustering properties of produced particles in
high-energy $pp$ collisions as a first step to identify minijets.  We
develop an algorithm to find clusters by using the k-means clustering
method, in conjunction with a k-number (cluster number) selection
principle in the space of pseudorapidity and azimuthal angles.  We
test the clustering algorithm using events generated by PYTHIA 8.1, for
$pp$ collision at $\sqrt{s}=200$ GeV. We find that clustering of
low-$p_T$ hadrons occurs in high multiplicity events.  However similar
clustering properties are also present for particles produced randomly
in a finite pseudorapidity and azimuthal angle space.  To distinguish
the dynamics from random generations of events, it is necessary to
examine the correlation between particles and between clusters.  We
find that the correlations between clusters may provide a useful tool
to distinguish the underlying dynamics of the reaction mechanism.

\end{abstract}


\maketitle
\section{Introduction}
\label{intro}

The mechanism of relativistic parton-parton hard scattering is an
important basic perturbative QCD particle production process in
high-energy nucleon-nucleon collisions
\cite{Bla74,Ang78,Fey78,Owe78,Rak13,Sjo86,Sjo87,Sjo05a,Sjo05b,Sjo06,Sjo07,Cor11a,Cor11b,Sjo14,UA1,Esk89,Wan91,Cal90,Cal90a,Cal91,Cal94,Cal94a,Cal01,Cal01a,Won94,Won12,Won13,Won15web,Won15}.
Because of the composite nature of a nucleon, multiple hard scattering
between partons of the projectile and target nucleons will lead to the
production of jets and dijets whose subsequent fragmentation gives
rise to the production of particle clusters.  It is different from the
nonperturbative flux-tube fragmentation process
\cite{Nam70,Bjo73,Cas74,Sch62,Art74,And79,And83,Art84,And83a,Sjo86,Sjo14,Sch51,Wan88,Pav91,Won91a,Won91b,Gat92,Won95,Feo08,Won94}
in which a quark of one nucleon and the diquark of the other nucleon
(or a gluon of one nucleon and the gluon of the other nucleon
\cite{Mc94,Mc01,Mc10a,Mc10b,Kov95,Gei09}) form one flux tube and the
subsequent fragmentation of the flux tube leads to the production of
hadrons.  It is also different from the direct-fragmentation process
\cite{Won78} in which the partons from the composite nucleon fragment
directly into the detected particles.

The hard-scattering process was originally proposed as the dominant
process for the production of high-$p_T$ jet clusters of order many
tens of GeV/c \cite{Bla74,Ang78,Fey78,Owe78,Rak13,Sjo86,Sjo87}.
However, the UA1 Collaboration found that it is also the dominant
process for the production of particle clusters with a total $p_T$ of
a few GeV/c for $p\bar p$ collisions at $\sqrt{s}$=0.2 to 0.9 TeV
\cite{UA1}.  The term ``minijet" was introduced to describe low-$p_T$
jet clusters \cite{Esk89}.  The dominance of jet production was found
to extend to lower $p_T$ domains at high collision energies because
(i) the fraction of particles produced by such a process increases
rapidly with collision energies $\sqrt{s}$, and (ii) the
jet-production invariant cross section at midrapidity varies as an
inverse power of $p_T$ \cite{Esk89,Wan91,Sjo06, Won13,Won15,Wonfn1}.

Recently, the region of dominance of the hard-scattering process has
been found to extend to the production of hadrons even to the lower
$p_T$ region of a few tenths of a GeV/c
\cite{Won12,Won13,Won15web,Won15}.  An indirect piece of evidence
comes from the observation on the transverse momentum spectra of
produced hadrons: For the production of particles with $p_T$ within
the range from a few tenths of a GeV to a few hundred GeV in
high-energy $pp$ and $p\bar p$ collisions at $\sqrt{s}$= 0.9 to 7 TeV,
the hadron transverse spectra, whose magnitude spans over 14 decades
of magnitude, can be described by a simple Tsallis inverse-power-law
type distribution with only 3 degrees of freedom
\cite{Won12,Won13,Won15web,Won15}.  The simplicity of the power-law
type transverse spectra suggests that only a single mechanism, the
hard-scattering process, dominates over the extended $p_T$ domain.  An
additional piece of direct evidence comes from the jetlike structure
in the two-hadron angular $(\Delta \eta, \Delta\phi)$ correlation data
in a minimum-$p_T$-bias measurement of the STAR Collaboration in $pp$
collisions at $\sqrt{s}=200$ GeV
\cite{star05,STAR06twopar,Por05,Tra11}.  The momentum distributions of
hadrons associated with a hadron trigger of a few GeV/c in $pp$
collisions at the same energy exhibit a jetlike cluster structure
within a cone in a similar manner, as observed by the STAR
Collaboration
\cite{STAR2p05,STAR2p06,STAR2p07,STAR2p07a,STAR2p07b,Won07,Won08,Won09}
and the PHENIX Collaboration \cite{PHENtwopar,Won09}.

The extension of the dominance of the hard-scattering model to the
low-$p_T$ domain of a few tenths of GeV/c raises serious questions on
the large and divergent perturbative quantum chromodynamics corrections at low $p_T$ and the
competition from nonperturbative flux tube fragmentation process
associated with low-$p_T$ phenomena. We need additional theoretical
and experimental comparisons of the hard-scattering model to construct
the proper phenomenological description in the low-$p_T$ region.

\begin{figure}[t]
\centering
\includegraphics[scale=1.00]{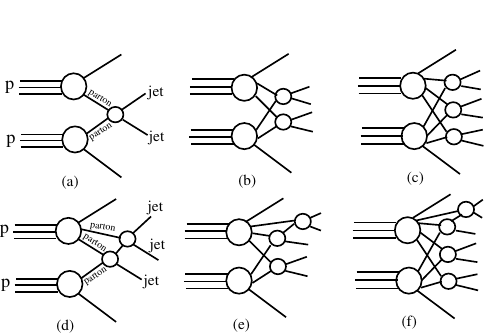}
\vspace*{0.3cm}
\caption{ Various multiple collision diagrams in a $pp$
  hard scattering leading to the production of jets, which are called
  minijets when the transverse momentum of the jet is small.  Shown
  here are diagrams for the production of (a) a dijet pair, (b) two dijet pairs,
  and (c) three dijet pairs. Furthermore, a scattered parton can make an
  additional collision with a different parton of the other proton, as
  shown in diagrams (d), (e), and (f).  }
\end{figure}

If the hard-scattering process of the $pp$ collision is appropriate
also for the low-$p_T$ region, then multiple parton interactions (known also as multiple collision processes)
\cite{Bla74,Sjo87,Sjo05a,Sjo05b,Cor11a,Cor11b} must occur to produce
multiple minijets and mini-dijets
\cite{Bla74,Sjo87,Sjo05a,Sjo05b,Sjo06,Cor11a,Cor11b,UA1,Esk89,Wan91}.
Among many other diagrams, the hard-scattering process can lead to the
production of one, two, and three pairs of mini-dijets as depicted in Figs.\ 1(a)-1(c).  Furthermore, a parton of one proton can make multiple
collisions (known also as rescattering \cite{Cor11b}) with different
partons of the other proton, as depicted in Figs.\ 1(d)-1(f).
The numbers of produced minijets can be even, as in Figs.\ 1(a)-1(c), or odd, as in Figs.\ 1(d)-1(f).  There can also be
additional higher-order diagrams with the radiation and the absorption
of gluon partons, which lead to additional minijets.

The multiple parton scattering processes in the production of
high-$p_T$ jets have been observed in high-energy $pp$ or $p\bar p$
collisions \cite{Ake87,Abe97,Aba10,Cha14}.  Theoretical discussions on
the production of minijets beyond the leading order has been
investigated, and hard inclusive dijet production with multiparton
interactions has also been considered
\cite{Cal90,Cal90a,Cal91,Cal94,Cal94a,Cal01,Cal01a,Won13,Kot17}.
However, in the low-$p_T$ region, the experimental investigation for
multiple parton interactions with the production of multiple minijets
and mini-dijets remains lacking.

We would like to develop tools to study multiple hard-scattering
processes for the production of multiple minijets and mini-dijets in
the low-$p_T$ domain in $pp$ collisions at high energies.  As a first
step, we examine here the clustering properties of minijets in the
pseudorapidity and azimuthal angle space and search for an algorithm
to assist the finding of minijet clusters candidates.

The minijet processes in a nucleon-nucleon collision are not only
intrinsically important with regard to our understanding of the
underlying mechanism for low-$p_T$ particle production, they are also
extrinsically valuable in applications because nucleon-nucleon
collisions lie at the heart of a nucleus-nucleus collision, and the
low-$p_T$ particle production dominates the particle production
process.  An understanding of the mechanism of low-$p_T$ particle
production in nucleon-nucleon collisions provide vital information on
the initial condition that may exist at the early stage of
nucleus-nucleus collisions, on which much interest has been focused
recently.  In particular, the observation of the near-side jet and the
away-side ridge in high-multiplicity events in high-energy $pp$
collisions
\cite{STAR2p05,STAR2p06,STAR2p07,STAR2p07a,STAR2p07b,Won07,Won08,Won09,Won11,CMS10,CMS16a,CMS16b,ATLAS16a,ATLAS16b,ALICE16a,ALICE16b},
indicates that the initial dynamics of the system after the production
of a jet or a minijet \cite{Won11} depends on the initial
configuration of the system. The examination of such a system also
calls for an event-by-event study of the multiple minijet and
mini-dijet productions in $pp$ collisions.

Our event-by-event study has been stimulated by a similar
investigation for particle production at lower $pp$ collision energies
where the particle production process may be dominated by flux-tube
fragmentation \cite{Won17}.  There, the basic conservation laws and
the semiclassical picture of the fragmentation process provide
powerful tools to reconstruct the space-time dynamics of the pair
production processes that may occur, if exclusive data for the
production process are available.  In the present investigation, the
space-time dynamics of a parton-parton hard scattering may provide
useful experimental information on the multiple collision processes
and on the constituent nature of the colliding nucleons.

In the search for separated minijet and mini-dijets, one of the
important ingredients is the $p_T$ threshold value that sets the $p_T$
limit for the inclusion of a particle as part of a minijet.  Clearly,
the higher the $p_T$ limit, the cleaner will be the cluster and their
possible corresponding minijet partners.  On the other hand, the
higher the $p_T$ value, the lower will be the number of cluster counts
and the lower the sampling statistics. Furthermore, because each
minijet occupies a substantial area in $(\eta, \phi)$ space, the
limited angular and azimuthal space may make the separation of the
minijets a more difficult task.  In the present manuscript, we shall
use the minimum-bias selection of particles with $p_T\ge 0.15$ GeV/c.
An optimum $p_T$ limit and cluster multiplicity will need to be
searched for in realistic applications with real data.

This paper is organized as follows.  In Sec. II, we summarize the
properties of a minijet from previous studies.  In Sec. III, we
exhibit the distribution of produced charged hadrons in the whole
range of rapidity and azimuthal angles for sample minimum-biased
PYTHIA calculations to illustrate the occurrence of clusters for $pp$
collisions at $\sqrt{s_{pp}}=200$ GeV.  In Sec. IV, we introduce
the algorithm for finding clusters in the pseudorapidity and azimuthal
angle space.  The algorithm consists of the k-means clustering method
supplemented by the k-number (cluster number) selection principle,
based on the physical properties of minijet clusters.  We illustrate
the usage of such an algorithm in Sec. V, using sample events with
high multiplicities generated by PYTHIA 8.1.  We examine the change of
the clustering behavior as a function of increasing multiplicities in
PYTHIA 8.1 events in Sec. VI.  We investigate whether similar
properties of clustering can be found in a random distribution within
the same finite $(\eta,\phi)$ phase space in Sec. VII.  We study
the correlation between particles and between clusters in Sec. VIII.  We present our conclusions and discussions in Sec. IX.  We
discuss another method of finding the cluster number, the elbow
method, and note its ambiguities in the Appendix.  For completeness,
we also include the results of the azimuthal angular correlations and
pseudorapidity correlations in the Appendix.

\section{ Properties of a Minijet }

The structure of a minijet in the $(\eta,\phi)$ scatter plot can be
inferred from the distribution of the two-hadron angular correlation
as a function of the pseudorapidity difference $\Delta
\eta$=$\eta_2-\eta_1$ and the azimuthal angular differences $\Delta
\phi$=$\phi_2-\phi_1$ of the two particles detected with angular
coordinates $(\eta_1,\phi_1)$ and $(\eta_2,\phi_2)$ in coincidence
\cite{star05,STAR06twopar,Por05,Tra11,STAR2p05,STAR2p06,STAR2p07,STAR2p07a,STAR2p07b,Won07,Won08,Won09,PHENtwopar}.
For $pp$ collisions at $\sqrt{s}=200$ GeV, the minijet structure
appears as a cluster of particles in the $(\eta,\phi)$ space (and a
cone in three-dimensional configuration space) as indicated by a
two-hadron Gaussian distribution in $\Delta \eta$ and $\Delta \phi$ in
the form
\begin{eqnarray}
\frac{dN}{d\Delta \eta ~d \Delta \phi}( \Delta \eta ,\Delta \phi) 
\propto \exp\left \{ - \frac{ (\Delta \eta)^2+\Delta \phi)^2}{2\sigma_\phi^2} \right \},
\label{eq1}
\end{eqnarray}
where the quantity $\sigma_\phi$ was found to be \cite{Won09} 
\begin{eqnarray}
\sigma_\phi =  \frac{ \sigma_{\phi 0}~m_a}{\sqrt{m_a^2+p_{T,\rm trigger}^2}}, ~~ \sigma_{\phi 0}=0.5,~ m_a=1.1 ~{\rm GeV},~~~
\label{eq2}
\end{eqnarray}
when triggered by a hadron with transverse momentum $p_{T,\rm
  trigger}$.    
It should, however, be emphasized that the Gaussian form of the
    distribution in Eq.\ (\ref{eq1}) is only a hypothesis.  Actual
    shape of the distribution will require the identification and the
    knowledge of minijets and all their individual member particles,
    which are not yet generally available.
In the minimum-bias data at the Relativistic Heavy Ion Collider energies, we shall
consider the quantity $p_{T,\rm trigger}$ takes on the value of
$\sqrt{ \langle p_T^2\rangle}$, which is of order 0.4 GeV/c.  Equation
\ (\ref{eq2}), therefore, yields $\sigma_\phi$$\simeq$0.5.  The
two-particle distribution of Eq.\ (\ref{eq1}) has a half width at half
maximum at $R$=$\sqrt{(\Delta \eta)^2+(\Delta \phi)^2}$=1.2$
\sigma_\phi$=0.6.
We can consider a circle of radius $R$ in the $(\eta, \phi)$ plane.
The minimum separation between any two points inside the circle is
zero and the maximum separation is $2R$.  Setting
$2R$=2.4$\sigma_\phi$ (or $R$=0.6) will allow the circle to contain a
large fraction (about 95\%) of the Gaussian distribution (\ref{eq1})
within the circular domain.  It is reasonable to assume that a
signature of a minijet cluster of particles is indicated by a cluster
of particles within a radius of $R$$\simeq$0.6 in the plane of
$(\eta,\phi)$.

In the hard-scattering process in the collision of two partons, $a+b
\to a'+b'$, the partons $a'$ and $b'$ materialize subsequently as
minijets.  The initial $a$ and $ b$ partons may be endowed with a
small intrinsic transverse momentum $k_T$ of the order of 0.6 to 1.0
GeV/c \cite{Kap78,Fey78,Owe84,Won98}.  The conservation of 4-momentum
requires that the scattered partons $a'$ and $b'$ will come out
azimuthally in nearly back-to-back directions.  The signature of a
mini-dijet can be taken to be a pair of minjets whose azimuthal
angles are approximately correlated within the range of $\pi-R$ to
$\pi + R$.

\section{Distributions of Produced Hadrons in Sample PYTHIA Events}

The description in terms of partons is useful only in the early stages
of the $pp$ collision. Subsequent evolution of the partons will
require their hadronization into detectable hadrons.  The dynamics of
particle production processes leaves an imprint on the distribution of
the produced particles.

Our knowledge of how partons hadronize remains incomplete.  We wish to
obtain some insight on the dynamics of the hadronization processes by
examining the distribution of the produced particles on an
event-by-event basis.  To see what may be expected, it is instructive
to study the distribution of produced particles in the PYTHIA 6.4
calculations with its hadronization model in which the history of the
evolution of the partons are recorded and traceable \cite{Sjo86}.

\begin{figure}[h]
\vspace*{-0.2cm} 
  \centering
  \includegraphics[scale=0.55]{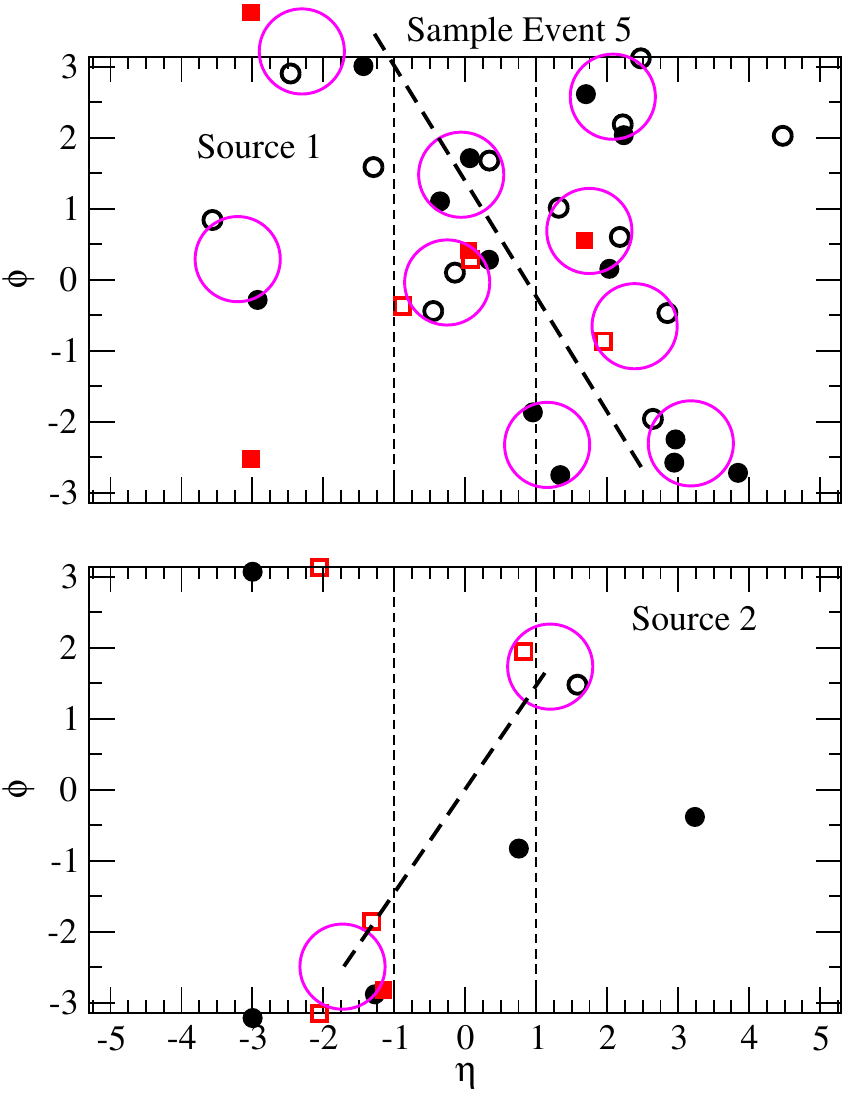}
\vspace*{-0.2cm}
\caption{ Scatter plots in the $(\eta,\phi)$ plane for
  produced charged particles in the full $(\eta,\phi)$ plane in a
  randomly selected sample event 5 generated by PYTHIA 6.4 for $pp$
  collisions at $\sqrt{s}$=200 GeV.  Cluster circles with a radius
  $R=0.6$ are plotted to circumscribe the cluster centers.  Some of
  the data points are wrapped around to facilitate cluster
  association.  }
\label{fig2n}
\end{figure}

In the PYTHIA 6.4 description of the $pp$ collision \cite{Sjo86},
valence quarks, valence diquarks, and gluon partons are produced and
they can be arranged into two initial strings connected by leading
valence quarks and antiquarks (or diquarks).  The produced gluons are
then split into quark-antiquark pairs and the quarks and their
neighboring antiquark (or diquark) are connected into segments of
shorter ``kinky substrings".  Each substring is subsequently
fragmented to produce quark-antiquark pairs in accordance with the
nonperturbative Lund string fragmentation model.  In the Lund model,
the fragmentation of the substring segments follows the
outside-inside cascades by producing a quark-antiquark pair carrying a
light-cone momentum fraction in accordance with a given fragmentation
function.  The $q$-$\bar q$ pair production leads to a shorter
remainder string with a smaller invariance mass, and the end parton
particles continue to repeat the string fragmentation process until
the invariance mass of the remainder string becomes lower than the
limit.  After the fragmentation of the "kinky" substrings,
neighboring $q$ and $\bar q$ (or diquark) are then connected to form
hadrons.  The production of the $q$-$\bar q$ pairs leads to clusters
of hadrons that are likely to be correlated at the end points and
along the string.  The outside-inside cascade of string fragmentation
of the leading partons of the string in the Lund model is
mathematically and kinematically similar to the parton cascade in high-$p_T$ leading parton fragmentation and parton showering, differing
mainly in the nature of the fragmentation functions.

\begin{figure}[h]
\vspace*{0.4cm} 
  \centering
  \includegraphics[scale=0.55]{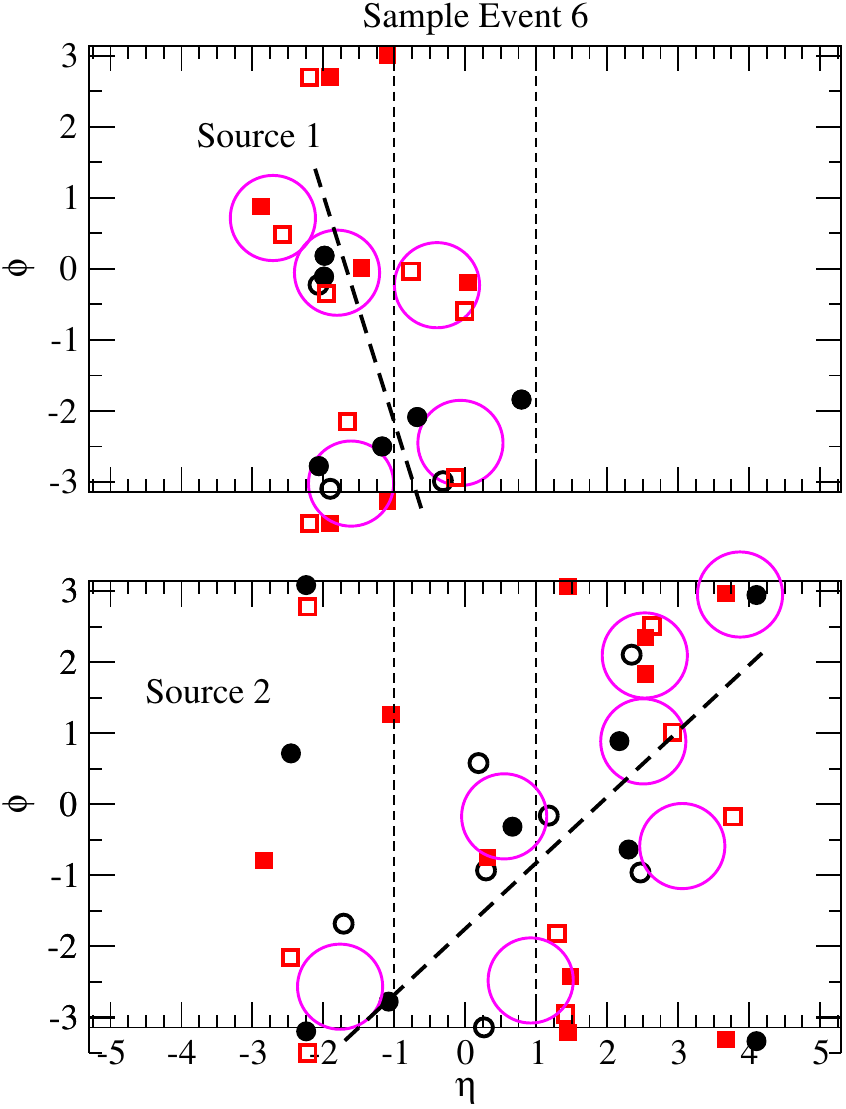}
\vspace*{-0.2cm}
\caption{  Scatter plots in the $(\eta,\phi)$ plane for
  produced charged particles in the full $(\eta,\phi)$ plane
 in  sample events   generated by PYTHIA 6.4
for $pp$ collisions at $\sqrt{s}$=200 GeV.
  Cluster circles 
  with a radius $R=0.6$ are plotted to circumscribe the cluster centers.
}
\label{fig3n}
\end{figure}

In a PYTHIA event, charged and neutral hadrons, as well as photons are
produced.  We shall focus on minimum-biased events without any $p_T$
selection.  They reside within the window of $y_{\rm target}\le \eta
\le y_{\rm beam}$, and $-\pi \le \phi \le \pi$ in the $(\eta, \phi)$
plane where $y_{\rm beam}$=$-y_{\rm target}$=5.29 for $pp$ collisions
at $\sqrt{s}=200$ GeV.

The scatter plots of produced charged hadrons obtained in PYTHIA 6.4
for a few randomly selected sample events, event 5 and event 6, are
displayed in Figs.\ \ref{fig2n} and \ref{fig3n}.  Each of the $pp$
events contains two separate quark-diquark sources of strings or
partons.  A string source will produce particles by string
fragmentation whereas a parton source will collide to produce
particles by the hard-scattering processes. To gain some insight on
the $p_T$ and the charge of the produced particles, we use circular
and square points to indicate $p_T$ less and greater than 0.5
GeV/c, respectively, with solid points for positive particles and open
points for negative particles.  In each event, the intermediate
outputs from PYTHIA 6.4 allow the specification of the two separate
strings or partons from which the produced charged hadrons originate.
The patterns of hadron particles reveal many interesting
characteristics.  One observes that produced particles tend to form
clusters.  A circle of radius $R=0.6$ and a minimum of two hadrons can
be conveniently used to separate different clusters, as such a
definition leaves very few numbers of hadrons outside the clusters.
For each string, the clusters appear correlated to form roughly a
linear pattern along the axis indicated by the dashed lines in
Figs.\ \ref{fig2n} and \ref{fig3n}.  Because of the fragmentation of
kinky substrings along the greater parent string, clusters have a
tendency to correlate with an azimuthal angular difference of about
$\pi$.  Source 1 of event 5 in Fig. 2 and source 2 of event 6 in
Fig.\ 3 give a large number of clusters along the parent string.  They
appear to bear the characteristic of a string fragmentation.  Source 2
of event 5 and source 1 of event 6 gives two groups of clusters, which
are roughly back-to-back correlated in the azimuthal degree of
freedom.  They appear to bear the signature of parton-parton
collisions.  Thus, both string fragmentation and parton-parton hard
scattering lead to clusters.  One expects intuitively that the string
fragmentation will likely lead to a chain of hadrons all along the
rapidity axis as in source 1 in event 5 and source 2 in event 6,
whereas a parton-parton hard scattering will lead to two groups of
clusters apart in rapidity, as in sources 2 in Event 5, and source 1
in event 6.

\section{  Algorithm for Finding   Clusters}

As discussed in Sec. II, a minijet shows up as a cluster of
hadrons with a cone radius of $R$=0.6 in the $(\eta,\phi)$ space.  The
sample events in PYTHIA 6.4 in Sec. III indicate that there are
clusters of produced particles in PYTHIA model calculations.  Minijets
are theoretically presumed to be produced in PYTHIA calculations.  It
is, therefore, useful to look for clusters as possible candidates for
minijets.

 Clusters, of the type shown in the last sections, can be searched for
 by the k-means clustering method
 \cite{Ng17,Ste57,Mac67,Lly82,Art07,Tho53,Che05}, in conjunction with
 an additional k-number (cluster-number) specification principle.  In
 such a search, we ascribe the characteristic cluster radius $R=0.6$
 to a cluster.  If the minijet producing hard-scattering process is
 dominant in the low-$p_T$ region, as suggested in earlier studies
 \cite{Won12,Won15,star05,STAR06twopar,Por05,Tra11}, then two clusters
 that are azimuthally correlated in a back-to-back manner have a high
 probability to be a physical mini-dijet of two correlated minijets
 at high collision energies.

For a given set of $M$ produced particles specified by their angular
positions $\{ {\bb x}_i$=$ (\eta_i,\phi_i)$,~$i$=1,2,3,...$M\}$ and
a given $K$ number of clusters, the k-means clustering method consists
of (i) partitioning the set of $M$ particles into $K$ cluster subsets
$S_k$=$ \{ {\bb x}_{i}^{k}\},~ k$=1,2,...,$K$ and (ii) finding for
each cluster subset the corresponding cluster center \break $\{{\bb
  C}_k$,~$ k$=1,2,...$K\}$ so as to minimize the potential function
\begin{eqnarray}
\Phi (K) =  \sum_{k=1}^K  \left \{ \sum_{{\bb x}_i^k \in S_k}  ({\bb x}_i^{k}-{\bb C}_k)^2\right \}  ,
\label{eq5}
\end{eqnarray}
which is defined as the total subset sum of the squares of the
distances between the cluster subset points and their corresponding
cluster center ${\bb C}_k$.

For a fixed value of $K$, the variation of the above potential
function $\Phi(K)$ with respect to the cluster center ${\bb C}_k$ is
given by
\begin{eqnarray}
\delta \Phi (K) = - \sum_{k=1}^K  \left \{ \sum_{{\bb x}_i^k \in S_k} 2({\bb x}_i^{k}-{\bb C}_k) \cdot \delta {\bb C}_k  \right \}  .
\end{eqnarray}
Because all $\delta {\bb C}_k$ are independent, the minimization of $\Phi(K)$
with respect to the variation of the positions of the cluster centers
${\bb C}_k$ leads to $\delta \Phi(K)/\delta {\bb C}_k$ = 0 and
 \begin{eqnarray}
 \sum_{{\bb x}_i^k \in S_k} 2({\bb x}_i^{k}-{\bb C}_k) =0.
\end{eqnarray}
This yields ${\bb C}_k$ as the centers of gravity of the subset of
points of $S_k= \{ {\bb x}_{i}^{k}\}, k=1,2,...,K$,
\begin{eqnarray}
 {\bb C}_k=\frac{1}{M_k}\sum_{{\bb x}_i^k \in S_k} {\bb x}_i^{k},
\label{eq6}
\end{eqnarray}
where $M_k=(\sum_{{\bb x}_i^k \in S_k}1)$ is the number (multiplicity)
of particles in the subset $S_k$.

In numerical implementation of the k-means clustering method for a
given value of the cluster number $K$, one chooses randomly the first
cluster center as one of the data points and chooses randomly the other
$K-1$ cluster centers in the other data points with probability
proportional to the square of the distance from the first cluster
center \cite{Art07}.  For each data point, the knowledge of the
positions of the initial cluster centers then allows one to calculate
the squares of the distance between the data point and all $K$ cluster
centers.  One then assigns each data point to the subset $S_k$ with
the smallest square of distance to its cluster center ${\bb C}_k$.
After all subset assignments to $S_k$ have been completed for all data
points, the center of gravity of the data points in each new subset
$S_k$ is then recalculated to give the new cluster centers ${\bb
  C}_k$, with which the iterative procedure will proceed until it
converges.  One then calculates the potential function $\Phi(K)$ of
Eq. (\ref{eq5}) as the sum of squared distances.

The above standard procedure is then repeated with other random
initializations of the initial cluster centers.  After many cluster
center random initialization, corresponding convergent
solutions, and the potential functions $\Phi(K)$ have been obtained,
the proper solution for the case of a given value of $K$ can be found
and selected as the solution with the minimum value of the potential
function $\Phi(K)$.  For a given value of $K$, the k-means clustering
method then yields uniquely the cluster subsets of particles $S_k= \{
{\bb x}_{i}^{k}\}, k=1,2,...,K$ associated with each cluster and the
corresponding cluster center location ${\bb C}_k$.

The k-means clustering method needs an amendment to make it applicable
for cluster searches because the method will lead to poorly displaced
and inaccurate cluster centers, if isolated particle points that are
obviously not part of a cluster and quite far away from a cluster have
been included into the particle data set in the clustering algorithm.
The presence of these isolated particles is possible because the
cluster partners of these isolated particles may not be detected within
the narrow window of acceptance, and there may further be other sources
of particle production in addition to those from clusters.  We need to
use our knowledge on the structure of the minijet in Eq.\ (\ref{eq1})
to sieve out these isolated data points in the set of $M$ particles.
We calculate the distances between any data point and all other data
points in the $(\eta,\phi)$ plane.  The knowledge of these distances
allows us to exclude any data point whose minimum separation to all
other data points exceeds a distance $2R$, presumably the maximum
separation for two data points in a minijet. (If we allow a degree of
fuzziness in excluding these isolated points, the minimum separation
can be set to $2R+2a$, where $a\ll R$ is the diffuseness parameter.)
After these points are excluded to yield a reduced set of particles
belonging to clusters in this modification, the k-means clustering
method becomes very efficient, fast converging, and capable of
yielding accurate cluster centers.  The method is stable against the
variations of the positions of the cluster centers, which turn out to
be the centers of gravity of the subset $S_k$ of the clustering
points, as given by Eq.\ (\ref{eq6}).  In this procedure, because the
azimuthal angle $\phi$ is equivalent to $\phi\pm 2\pi$ with a modulo
of $2\pi$, it is important to wrap around the azimuthal angles when
such a wrapping leads to an additional possibility of minijet
clustering.

We presume on the outset that a cluster consists of at least two
particles.  The k-means clustering method requires a prior knowledge
of the cluster number $K$. There may be different ways to partition a
group of $M$ particles into different numbers of clusters and the
locations of the cluster centers may also vary.  The selection of $K$
and the identification of particles as belonging to different $K$
clusters may, therefore, be ambiguous.  Our algorithm to find clusters
must contain an additional method to select the appropriate cluster
number $K$ that is based on well-founded physical principles.

For a given set of $M$ produced particles on the $(\eta,\phi)$ plane,
one considers a possible range of cluster $K$ numbers, $K=K_{\rm
  min},...,K_{\rm max}$.  The maximum limit $K_{\rm max}$ occurs when
the cluster number $K_{\rm max}$+1 leads to the forbidden case of
having a cluster with only a single particle.  For each cluster number
$K$ in the range under consideration, the k-means clustering method
leads to a unique partition into $K$ clusters with their corresponding
cluster centers ${\bb C}_k$.  To select the appropriate $K$, we use
the minijet physical properties discussed in the last section that a
cluster circle with a radius $R$=0.6 of a physical minijet contains
almost all of the particles of the physical minijet.  In order for
the cluster number $K$ to lead to the appropriate partition of the set
of $M$ particles into $K$ physical minijet or clusters, the
corresponding $K$ cluster circles with a radius $R=0.6$ should contain
all, or almost all, $M$ data points of the set.  There should be very
few data points outside the cluster circles.  The k-number (cluster
number) selection principle is, therefore, that {\it $K$ should be the
  cluster number that leads to the fewest number of data points $\Omega$ outside the cluster circles with an assumed radius}.

In the process of determining quantitatively the number of outside
data points, one finds that there are often some data points close to
the circular boundary, which can be considered as part of the cluster.
To account for such a possibility of inclusion of these hadrons into
the clusters, we generalize the number of outside points from a
discrete number $\Omega$ to a continuous quantity by
\begin{eqnarray}
\Omega=\sum_{k=1}^K\left \{
\sum_{{\bb x}_i^k \in S_k} 
\left  [1-\frac{1}{1+\exp\{\frac{|{\bf x}_i ^k- {\bf C}_k|-R}{a}\}}\right ]\right \}
\end{eqnarray}
where for our case, we have taken the value $R=0.6$.  In the case with
a sharp boundary $a\to 0$, we just have the case of a discrete number
of outside points.  We shall use {{\color{blue}}$a$=0.1 } for
numerical purposes.  In applying the principle of the least number of
outside points, we calculate the generalized $\Omega$ only for points
close to the cluster's boundary with the region between $R \sim
R+a$. We directly reject points beyond $R+a$ as they are too far away
from the clusters, and the possibility to involve these points inside
the clusters is also very low. What is more, we also directly involve
the points within $R$ inside the clusters. By these ways, we can make
the sharp circle clusters to be flexible, and we still ensure the
algorithm to be stable and fast.

For each iteration in each event, there may be particles farther away
from all cluster centers beyond the separations of $R+a$ after
particles are partitioned into sets of clusters.  These data points
will not be included in the determination of the new cluster centers
for the next iteration.

By generalizing the number of outside points $\Omega$, from a discrete
number to a continuous quantity, the principle of smallest outside
points choice of $K$ is such that $K$ is that the quantity
$\Omega$ is smallest for different $K$.  If there are two $K$ values
having the same fewest outside points within a range, we should select
the smaller $K$ value because the set of the smaller number of
minijets can radiate a parton and become the parent of the set with a
greater number of minijets.

In summary, our cluster finding algorithm, therefore, consists of the
k-means cluster method, supplemented by the k-number selection
principle of the fewest number of data points outside of the cluster
circles.

\section{Illustration of the Algorithm for finding     Clusters }

We shall apply the above algorithm for finding clusters from charged
hadrons generated by the PYTHIA 8.1 for high-energy $pp$ collisions
at~$\sqrt{s}$=200 GeV.  The event generators PYTHIA 8.1 \cite{Sjo07}
and PYTHIA 6.4 \cite{Sjo06} include the multiple parton interaction
processes as described in Ref.\ \cite{Sjo87}, with additional
considerations on color correlations, flavor correlations, junction
topology, beam remnant configurations \cite{Sjo05a}, and interleaving
initial state radiations \cite{Sjo05b}.  The fully interleaving
evolution \cite{Cor11a} and rescattering \cite{Cor11b} are further
included in PYTHIA 8.2 \cite{Sjo14}.

In the series of PYTHIA programs, the basic picture of the multiple
collision process arises from the composite nature of the proton which
possesses a parton spatial distribution in addition to the standard
parton momentum distribution (parton distribution function).  The parton-parton
collisions between the constituents of the projectile proton and the
target proton are assumed to be independent of each other, and the
number of collisions in an event is, therefore, given by a Poisson
distribution. The probability of parton-parton collisions is then a
function of the parton-parton cross section and the impact parameter.
To extend the parton-parton scattering cross section to the low-$p_T$
region for minimum-bias studies, the divergent parton-parton
scattering cross section at low transverse momenta has to be
regularized with a cut-off parameter that can be chosen to yield the
appropriate charged-hadron multiplicity distribution. We expect finite
multiple parton-parton multiple collision probabilities for the
independent collisions of projectile partons with target partons as
depicted in the diagrams in Fig. 1.  They lead to the production of
multiple minijets and mini-dijets in the angular scatter plots of
produced charged particles.

The probability for the occurrence of minijets and mini-dijets
depends on the charge multiplicity of the event, which is part of the
total hadron multiplicity.  For brevity of notation and its frequent
usage, we shall abbreviate ``charge multiplicity" or
``charged-particle multiplicity" simply by ``multiplicity" when
ambiguities do not arise or are not pertinent.  We can restore back
the term ``charge multiplicity" when it is properly needed.

In order to predict what may be expected experimentally for multiple
minijet and mini-dijet productions, we generate minimum-bias events
using the PYTHIA 8.1 and we accept primary charged particles with
$|\eta|$$\le$ 1.  For each event multiplicity, we select five random
events for illustration.  We shall label each event by the index
p$M$e$I$, where p$M$ stands for PYTHIA minimum-biased event with charge
multiplicity $M$, and e$I$ denotes {\it }event number $I$ with the
charge multiplicity $M$.  We would like to search for the presence of
the expected and mini-dijetlike clusters from the angular scatter
plots of charged particles in these events.

\begin{figure}[h]
\vspace*{0.4cm} 
  \centering
  \includegraphics[scale=1.00]{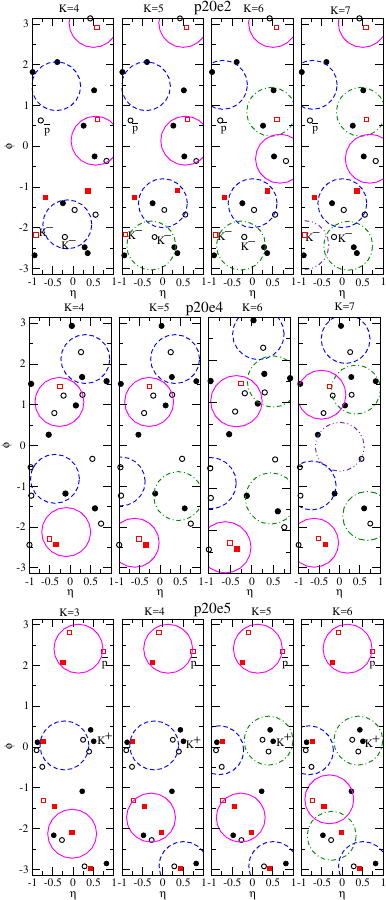}
\vspace*{-0.2cm}
\caption{ Scatter plots in the $(\eta,\phi)$ plane for
  produced charged particles in minimum-bias events with multiplicities $M$=20
for $-1$$\le$$\eta$$\le$1,
 in  sample events p20e2, p20e4, and p20e5,  generated by PYTHIA 8.1
for $pp$ collisions at $\sqrt{s}$=200 GeV.
  Cluster circles 
  with a radius $R=0.6$ circumscribe cluster centers
  obtained with the k-means clustering method
  assuming different cluster numbers $K$.  
}
\label{fig4n}
\end{figure}

The detected and identified charged particles include not only charged
hadrons but also a small percentage (of about 12\%) of $e^+$ or $e^-$.
By convention, we include these leptons in our charged multiplicity
counts.  However, because the $e^+$ and $e^-$ particles arise from
many different hadronic and nonhadronic sources, and the relations
between these particles and their hadron parents, if they arise from
hadronic decays, are nontrivial, we shall exclude them in our
minijet finding algorithm.  Their presence in the scattered
$(\eta,\phi)$ plot provides a sense of possible hadronic activities in
the vicinity of their angular locations.

In Figs.\ 4, 6, 7, and 8, we shall show sample scatter plots of
charged particles in the $(\eta, \phi)$ plane from minimum-bias events
simulated by the PYTHIA 8.1 event generator.  We display the particle
labels of kaons, protons, electrons, and muons while the other
particles are all charged pions.  The solid and open points denote
positive and negative particles respectively, and circular and square
points denote $p_T\le 0.5$ GeV/c and $p_T > 0.5$ GeV/c, respectively.

We shall illustrate the algorithm for finding clusters with concrete
examples.  We consider three randomly selected minimum-bias PYTHIA 8.1
events with $M$=20 in Fig.\ \ref{fig4n}.  For each of these events, we
assume different cluster numbers $K$ and obtain $K$ clusters and
their corresponding cluster centers ${\bb C}_k$ using the k-means
clustering method.  We then construct cluster circles with a radius
$R=0.6$ circumscribing the cluster centers.

In Fig.\ \ref{fig4n}, for Event p20e2 with $K=$ 4, 5, 6, and 7 on the
top panel, the number of points $\Omega$ outside of the cluster
circles are 10, 6, 4, and 2, respectively.  For the case of $K$=8,
there is no k-means clustering solution without one of the clusters
possessing only a single particle.  Because we do not consider a
single particle to be a cluster, $K$=8 is excluded from our
consideration for event p20e2.  If the clusters are minijet clusters,
then almost all particle points should be inside the cluster circles.
The case of $K=7$ leads to the fewest number of particles $\Omega$
outside of the cluster circles.  According to the principle of fewest
outside points, $K=$7 is the proper number of clusters for event p20e2
on the top panel.  Similarly, for event p20e4 in Fig.\ \ref{fig4n}
with $K=$ 4, 5, 6, and 7 in the middle panel, the number of points
outside of the cluster circles are 8, 5, 3, and 0, respectively.  We
infer that $K=$7 leads to clusters for event p20e4.  For event p20e5
in Fig.\ \ref{fig4n} with $K$=3, 4, 5, and 6 in the lower panel, the
number of outside points are 11, 6, 1, and 0.  We infer that $K=6$ is
the proper cluster number with zero points outside of the cluster
circles.

It should be mentioned that there is another method, the ``elbow
method", to select the cluster number $K$ by studying the
$K$-dependence of the potential function $\Phi(K)$ \cite{Ng17,Tho53}.
The method consists of determining the cluster number by the location
of the ``kink" where there is a sudden change of the slope of the
potential function.  The method suffers from the ambiguities in
finding where the kink lies, and will not be used in the present
context.  We shall discuss the ambiguities in such a method in
Appendix A.

\section{Scatter Plots of Produced Charged Particles from PYTHIA 8.1}

\begin{figure}[h]
\hspace*{-0.2cm}
\centering
\includegraphics[scale=0.43]{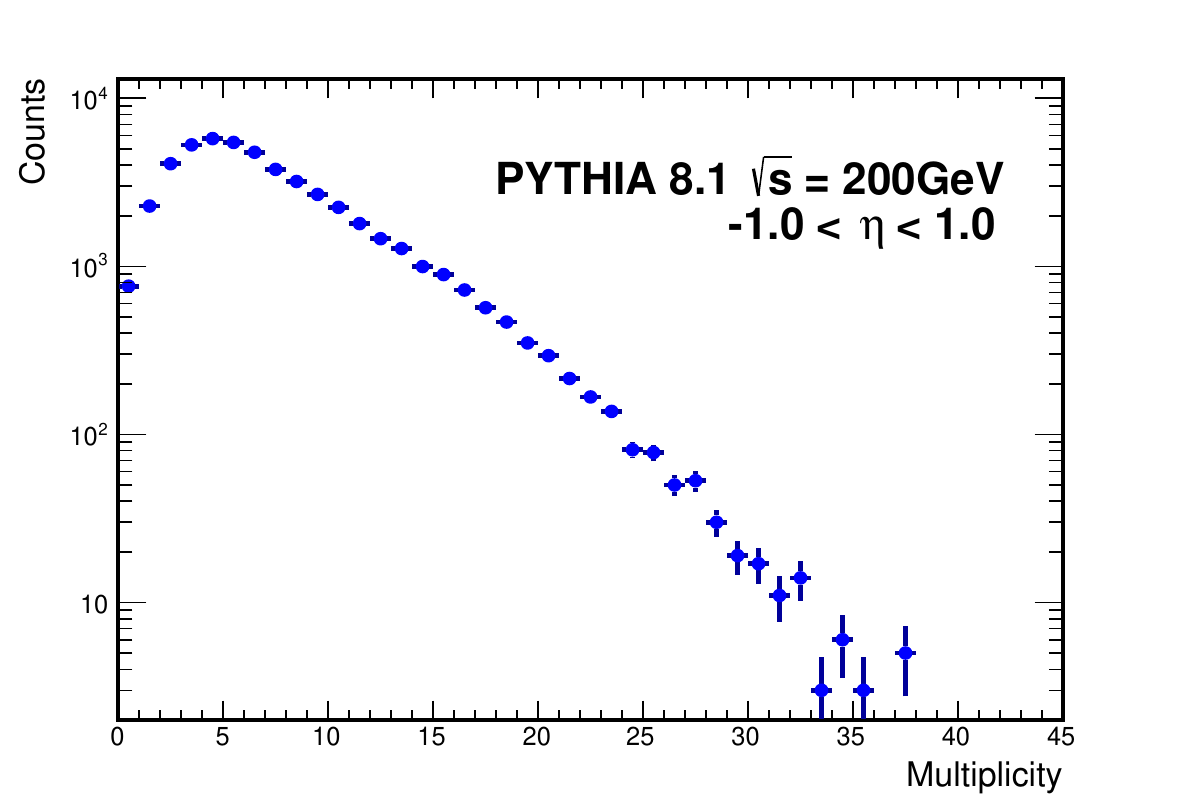}
\vspace*{-0.0cm}
\caption{Multiplicity distribution $dN/dM$ of charged particles
  within the window $-1$$\le$$\eta$$\le$1 obtained with PYTHIA 8.1, for $pp$
  collisions at $\sqrt{s}=200$ GeV.}
\label{fig5n}
\end{figure}

We study the clustering properties of charged particles produced in
$pp$ collisions in events with $-1 \le \eta \le 1$, $-\pi \le \phi \le
\pi$ and generated by PYTHIA 8.1 at $\sqrt{s}=200$ GeV without a $p_T$
selection.  In reviewing the scatter plots in the $(\eta,\phi)$ space
as a function of the charged particle multiplicity, it should be kept
in mind that those events with larger charge multiplicity numbers $M$
are events with lower occurrence frequencies as given in
Fig. \ref{fig5n}.  The average number of charged particles within the
window of $|\eta| \le 1$ is $\langle M$$ \rangle $=6.94.

\begin{figure}[t]
  \centering
 \includegraphics[scale=1.00]{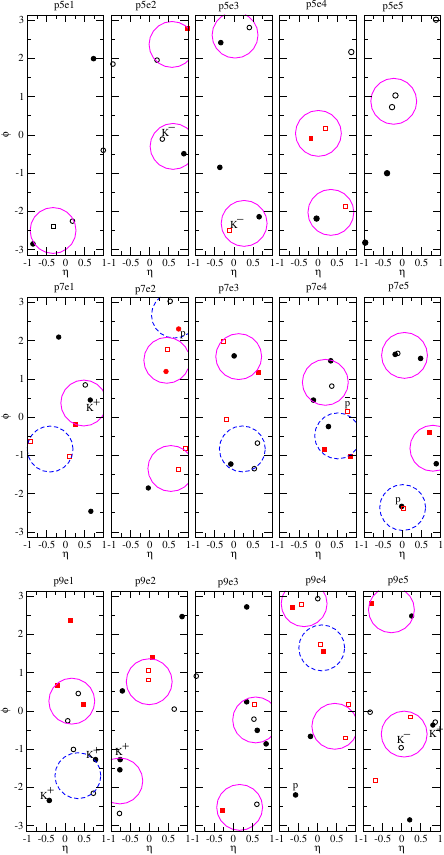}
\caption{(Color Online) Scatter plots in the $(\eta,\phi)$ plane for
  produced charged particles in events, with multiplicities $M$=5, 7,
  and 9, within $-1$$\le$$\eta$$\le$1, generated by the PYTHIA8.1 for
  $pp$ collisions at $\sqrt{s}$=200 GeV.  Circular curves indicate the
  locations of the clusters.  }
\label{fig6n}
\end{figure}

We plot in Figs.\ \ref{fig6n}~\ref{fig8n} clusters of particles
within a radius of $R=0.6$ obtained from the clustering algorithm.  As
the multiplicity increases beyond $M$=6, there appears to be a gradual
onset of the production of multiple clusters for $pp$ collision at
$\sqrt{s}=200$ GeV.

An interesting question arises whether the angular clustering of data
points at $(\Delta \eta, \Delta \phi)\sim$ 0 may arise from the decay
of resonances.  For a resonance with a mass $M$ decaying into two
particles with momenta $p_i$=$(y,p_{Ti},\phi_i)$ with $i$=1,2 and
transverse masses $m_{Ti}$=$\sqrt{p_{Ti}^2+ m_i^2 }$, the angular
correlation of the two particles for $\Delta y$=$(y_1-y_2)$ and
$\Delta \phi$=$\phi_1-\phi_2$ satisfies
\begin{eqnarray}
&&\hspace*{-0.6cm}\frac{M^2-m_1^2-m_2^2}{2} 
\nonumber\\
&&~~~~=m_{T1} m_{T2}\cosh (\Delta y)   - p_{T1}p_{T2}\cos(\Delta \phi).
\end{eqnarray}
For small $|\Delta y|$ and $|\Delta \phi|$, we can expand the $\cosh$
and $\cos$ functions and get
\begin{eqnarray}
&&m_{T1}m_{T2} (\Delta y)^2 +p_{T1}p_{T2} (\Delta \phi)^2
\nonumber\\
&& ~~~~=M^2-m_1^2-m_2^2-2m_{T1} m_{T2}+2 p_{T1}p_{T2}.
\end{eqnarray}
In the decay into two masses, the scatter plot of the two final
particles for small values of $\Delta y$ and $\Delta \phi$ fall within
an ellipse with ellipsoidal radii given by
\begin{subequations}
\begin{eqnarray}
a_{\Delta y}=\sqrt{\frac{M^2-m_1^2-m_2^2-2(   m_{T1}m_{T2} -p_{T1}p_{T2} )}{m_{T1}m_{T2}}},~~~~~
\\
a_{\Delta \phi}=\sqrt{\frac{M^2-m_1^2-m_2^2-2(   m_{T1}m_{T2} -p_{T1}p_{T2} )}{p_{T1}p_{T2}}}.~~~~~~
\end{eqnarray}
\label{radii}
\end{subequations}
\begin{figure}[t]
 \centering
\includegraphics[scale=1.00]{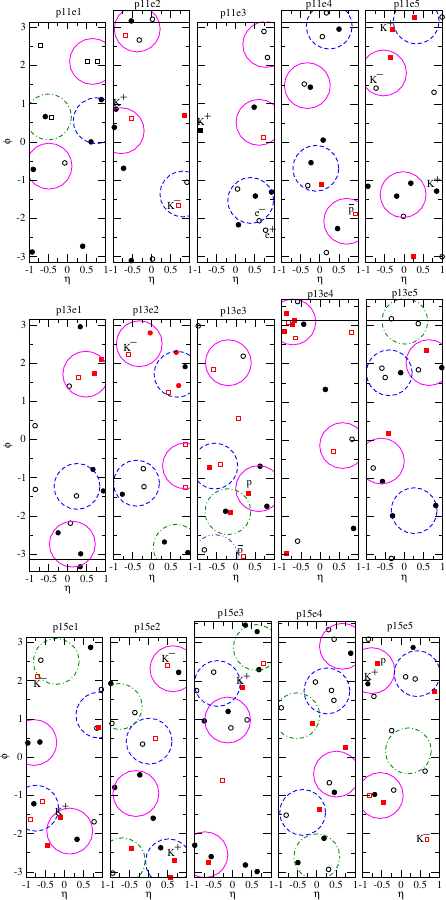}
\vspace*{0.2cm} 
\caption{(Color Online) Scatter plots in the $(\eta,\phi)$ plane for
  $M$=11, 13, and 15, within the window of $-1$$\le$$\eta$$\le$1, generated by
  the PYTHIA8.1 for $pp$ collisions at $\sqrt{s}$=200 GeV.  Circular
  curves indicate the locations of the   clusters.  }
\label{fig7n}
\end{figure}
Thus, the decay of a resonance may appear as a cluster within a radius
$a_{\Delta y}$ and $a_{\Delta \phi}$ and not necessarily and directly
from a minijet, depending on the quantities as given on the right-hand side of the above equations.  Upon approximating the rapidity $y$
as the pseudorapidity $\eta$, the above results show that the decay of
a resonance may appear as a cluster with the radii of
Eq.\ (\ref{radii}).

The partitioning of the set of charged particles into clusters can be
carried out on an event-by-event basis in Figs.\ \ref{fig6n}~\ref{fig8n} by identifying a cluster as an assemble of particles represented by a circle in the $(\eta,\phi)$ plane with a radius of
$R=0.6$. We can furthermore identify a mini-dijetlike pair of
clusters as two correlated clusters whose centers are separated
azimuthally within the range from $\pi - R$ to $\pi + R$.  In
Figs.\ \ref{fig6n}~\ref{fig8n}, we indicate a cluster and its
corresponding associated partner by circles of the same line type and
color.  At the end edges of $\phi=\pm \pi$, the scatter plots are
sometimes wrapped around so as to facilitate the partitioning
particles into clusters, as in events p11e2,p11e4,p11e5,...

The data in Figs.\ \ref{fig6n}~\ref{fig8n} reveal that as the
multiplicity increases, clusters of more than two particles within a
radius of $R$=0.6 occur with a greater probability.  In most of the
events with $M$=7 to 9 and higher multiplicities, a single cluster
appears often to correlate roughly with an associated partner in
azimuthally nearly back-to-back directions.  There may be a
fluctuation of the back-to-back correlation due to the intrinsic
transverse momentum of the partons.  We conclude from these figures
that mini-dijetlike clusters commence at $M$$\sim$7 with the
probability increasing gradually as $M$ increases and appear nearly
consistently for $M \gtrsim 11$, as indicated in Figs. 7 and 8.

We show in Fig.\ \ref{fig7n} the scatter plots of charged particles in
events with high multiplicities $11 \le M \le 15$.  As the
multiplicity number $M$ increases beyond $M \gtrsim 13$, there is a
transition from the production of one pair of mini-dijet-like clusters
to the production of two pairs of mini-dijetlike clusters, with each
pair of mini-dijetlike cluster approximately azimuthally back-to-back
with respect to each other.  The transition region is not sharp as
many events contain only a single pair of mini-dijetlike cluster,
while many other events in Fig. \ref{fig7n} contain double correlated
mini-dijetlike clusters.  We conclude from these figures that two
mini-dijetlike cluster pairs begin to set in with $M\gtrsim 14$ with
the probability increasing gradually as $M$ increases.

\begin{figure}[h]
  \centering
 \includegraphics[scale=1.00]{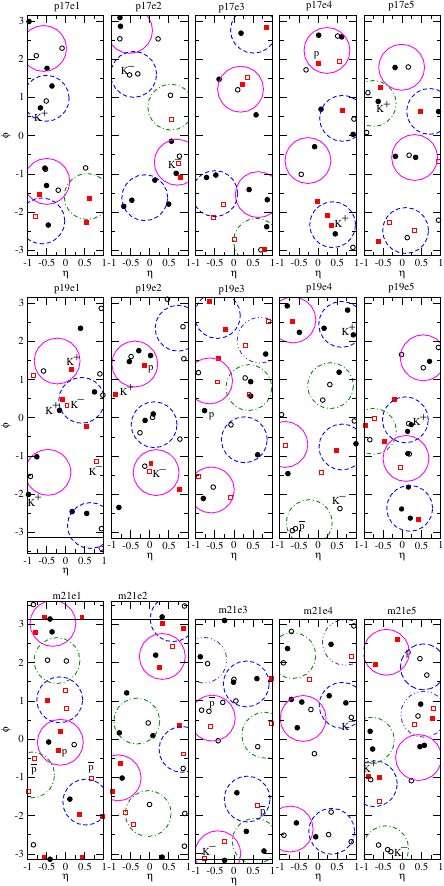}
\caption{ Scatter plots in the $(\eta,\phi)$ plane for
  $M$=17, 19, and 21 within the window of $-1$$<\eta<$1 generated by
  the PYTHIA 8.1 for $pp$ collisions at $\sqrt{s}$=200 GeV.  Circular
  curves indicate the locations of the clusters.  }
\label{fig8n}
\end{figure}

We show in Figs.\ \ref{fig8n} the scatter plots of charged particles
in events with ultra-high multiplicities $17 \le M\le$21.  As the
multiplicity number $M$ increases beyond $M\gtrsim 17$, the production
of two sets of mini-dijetlike clusters appears nearly consistently
with occasional production of five clusters.  In Fig.\ \ref{fig8n},
events with $M\gtrsim20$ appear to contain events with three pairs of
mini-dijetike clusters.

The results from the present analysis indicates that multiple clusters
and mini-dijetlike clusters are common occurrences for events with
high multiplicities and their numbers increase with the increasing
multiplicity $M$.
 
\begin{figure}[b]
 \hspace*{0.5cm}
\includegraphics[scale=0.40]{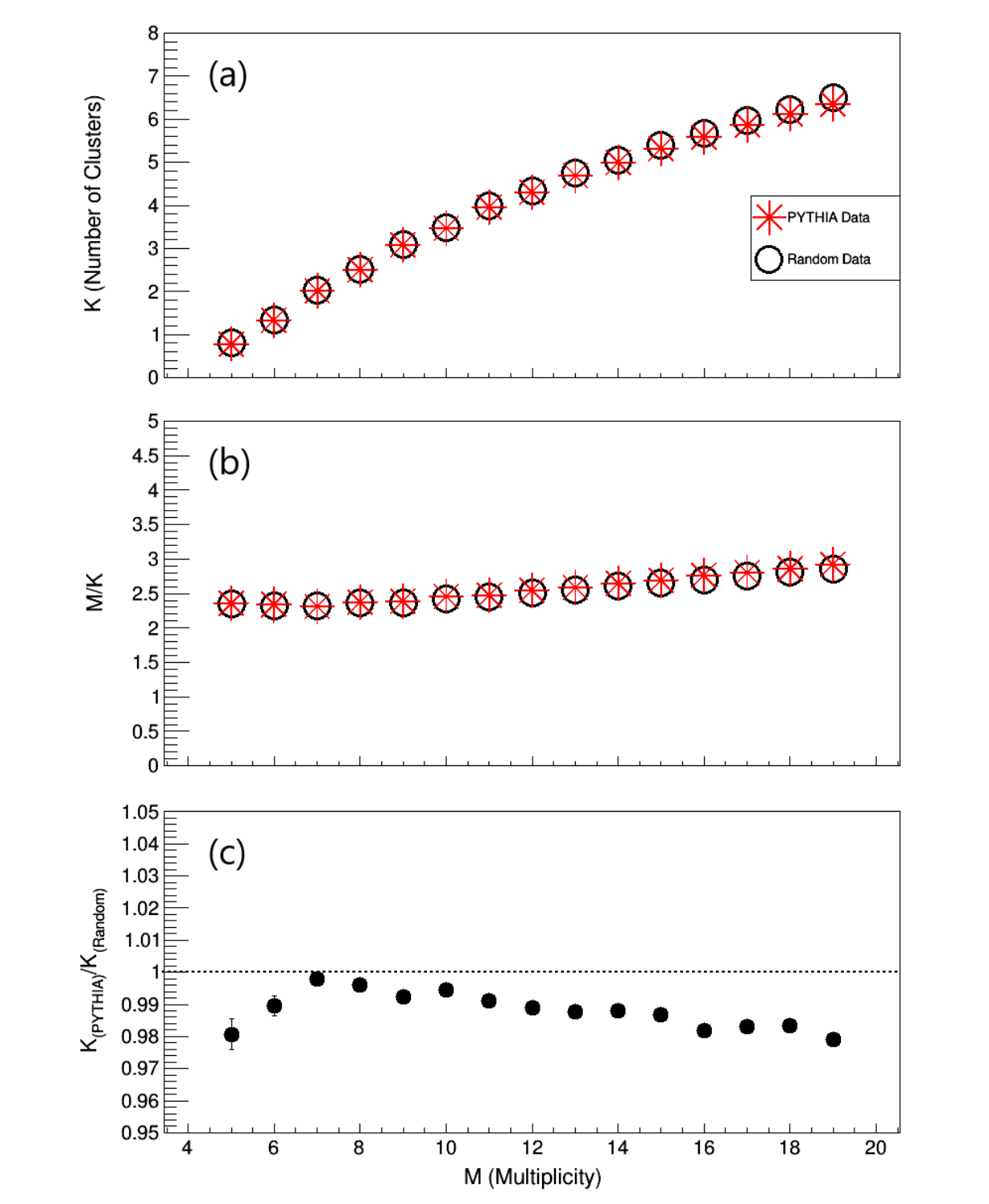}
\vspace*{0.3cm} 
\caption{(Relations between charge multiplicity $M$ and
  the number of clusters $K$: for $pp$ collisions at $\sqrt{s}=200$
  GeV as extracted from events generated by PYTHIA 8.1 within
  $|\eta|\le 1$ and $|\phi|$$\le$$ \pi$.  (a) The relation of $K$ as a
  function of $M$ obtained from PYTHIA (star symbol) and the random
  distribution (open circle).  (b) The ratio of $K/M$ as a function of
  $M$ for particles.  (c) The ratio of $K_{\rm PYTHIA}/K_{\rm random}$
  as a function of $M$.}
\label{fig9n}
\end{figure}

Figure\ \ref{fig9n}(a) shows that for events generated by PYTHIA 8.1
within $|\eta|\le$ 1, the number of clusters $K$
appears to increase monotonically and approximately as a linear
function of charge multiplicity $M$.  The relationship between
$(M/K)_{\rm PYTHIA}$ and $M$ is shown in Fig \ref{fig9n}(b).  The
ratio $(M/K)_{\rm PYTHIA}$ is 2.355
for M=5 and is 2.372
for M=8.

\section{ Clustering of Particles in a Random Distribution}

The results in the last section indicate the copious production of
clusters in the theoretical model of PYTHIA 8.1.  Many of these
clusters also exhibit back-to-back azimuthal correlations to make them
good candidates for physical mini-dijets.  These theoretical clusters
as well as their corresponding experimental counterparts will likely
represent physical minijets and mini-dijets, if the dominance of the
parton-parton hard-scattering process for minijet production is
extended to the low-$p_T$ region, as suggested in
\cite{Won12,Won15,star05,STAR06twopar,Por05,Tra11}.

It is worth noting that the clustering property by itself is not
sufficient to definitively identify a cluster as minijet cluster
because similar clustering properties may also be present in other
particle production models.  It is necessary to have other independent
collaborative supports for the minijet occurrence in order to
identify the observed clusters as likely physical minijet clusters.

In order to bring the need for independent collaborative supports into sharp focus, it is illustrative to examine the clustering properties
of particles produced in a simple schematic model in which a total of
$M_{\rm random}$ number of particles are randomly and independently
produced with a uniform probability in the $(\eta,\phi)$ phase space
within the window of $| \eta|$$ \le $$\Delta \eta_{\rm window}/2$ and
$ |\phi|\le \pi $,
\begin{eqnarray}
\frac{dP_{\rm random}}{d \eta ~d\phi}=\frac{ \Theta ( \Delta \eta_{\rm window}/2-   |\eta|  )~~\Theta(\pi -|\phi|)}{2\pi \Delta \eta_{\rm window}}.
\label{eq8}
\end{eqnarray}
This can be the approximate mode of production when particles are
produced independently with a uniform probability in rapidity, as from
the fragmentation of a flux tube at very high energies
\cite{Nam70,Bjo73,Cas74,Sch62,Art74,And79,And83,Art84,And83a,Sjo86,Sjo14,Sch51,Wan88,Pav91,Won91a,Won91b,Gat92,Won95,Feo08,Won94}.
It can also be the probability distribution used to describe noise
particles randomly produced within the experimental $(\eta,\phi)$
phase space.  We use the symbols $\{M_{\rm random},K_{\rm random}\}$
to denote the multiplicity number and cluster numbers respectively,
using a random generation of particles.
 
We find in this case of random distribution that particle clustering
also occurs when a large $M_{\rm random}$ number of particles are
produced randomly over a small phase space.  To understand such a
clustering, we can pick any two produced particles.  The probability
that a pair of particles falling randomly within the circle of radius
$R$ with respect to each other is
\begin{eqnarray}
P_{\rm random}= \left( \frac{\pi R^2}{ \Delta \phi_{\rm window}\Delta \eta_{\rm window}}\right ),
\label{eq12n}
\end{eqnarray}
where $\Delta \phi_{\rm window}$=2$\pi$ and
$\Delta \eta_{\rm window}$=2 for the present window.
In an
event with multiplicity $M_{\rm random}$, the number of  distinct pairs is
\begin{eqnarray}
({\rm number ~of~ distinct~pairs}) \!= \!\frac{M_{\rm random}(M_{\rm random}-1)}{2}
\! .~~~~~~
\label{eq13n}
\end{eqnarray}•
Therefore, with multiplicity $M_{\rm random}$, the (average) number of
clusters, $K_{\rm random}(2,M_{\rm random})$, is the product of
Eqs.\ (\ref{eq12n}) and (\ref{eq13n}),
\begin{eqnarray}
K_{\rm random}(2,M_{\rm random}) =\frac{M_{\rm random}(M_{\rm random}-1)}{2
(2\pi \Delta \eta_{\rm window})}\pi R^2,~~~~~~
\label{eq14n}
\end{eqnarray}
upon identifying a cluster as two particles falling within a radius of
$R=0.6$.  However, because clusters can be formed with more than two
particles, the above quantity $K_{\rm random}(2,M_{\rm random})$
represents only the upper limit of the number of clusters when
particles fall into and join other clusters.

More generally, the number $K_{\rm random}(n,M_{\rm random})$ of
clusters of random coincidence for a cluster of $n$ particles within a
radius of $R$ in an event with multiplicity $M$ is
\begin{eqnarray}
K_{\rm random}(n,M_{\rm random})=C_n^{M_{\rm random}} \left( \frac{\pi R^2}{2\pi \Delta \eta_{\rm window}}\right )^{n-1}   \!\!\!\!\!\!\!\!.
  ~~~~~~\end{eqnarray}
 For a detector such as the STAR detector with a pseudorapidity window
 $\Delta \eta_{\rm window}$=2, we have
\begin{eqnarray}
K_{\rm random}(2,M_{\rm random}) \!=\!\frac{M_{\rm random}(M_{\rm random}\!\!-\!1)}{2}\!\! \times\! 0.09.~~~~~
\label{eq16n}
\end{eqnarray}

\begin{figure}[h]
  \centering
 \includegraphics[scale=1.00]{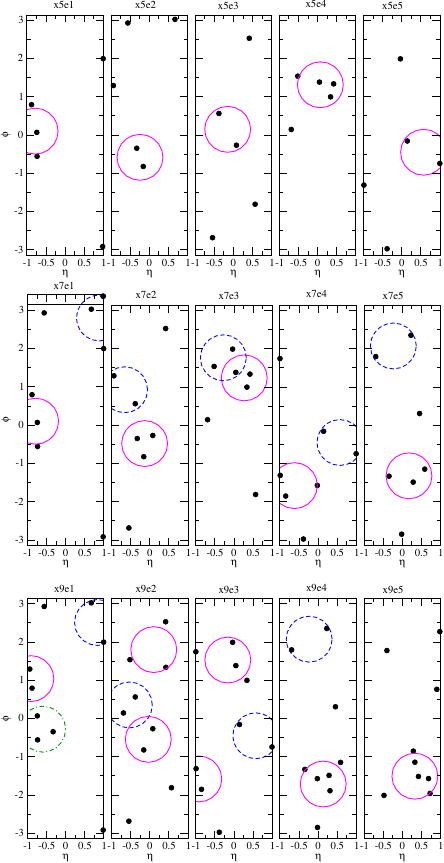}
\caption{Scatter plots in the $(\eta,\phi)$ plane for
  produced particles in events with multiplicities $M_{\rm random}$=5,
  7, and 9, production within $|\eta|\le 1$ and $|\phi|$$\le$$ \pi$.
  Circular curves indicate the locations of the cluster circles with
  $R=0.6$.  }
  \label{fig10n}
\end{figure}

\begin{figure}[h]
  \centering
\includegraphics[scale=1.00]{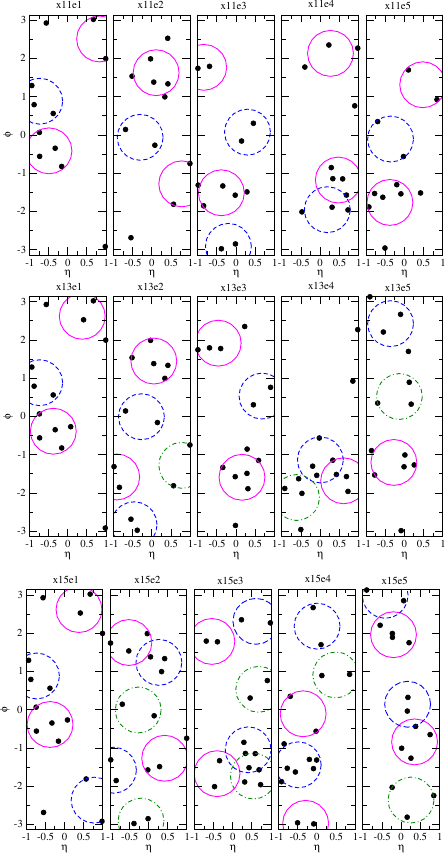}
\caption{Scatter plots in the $(\eta,\phi)$ plane for
  produced particles in events with multiplicities $M_{\rm
    random}$=11, 13, and 15 generated by an event generator with a
  uniform and independent production within $|\eta|\le 1$ and
  $|\phi|$$\le$$ \pi$.  Circular curves indicate the locations of the
  cluster circles with $R=0.6$.  }
  \label{fig11n}
\end{figure}

\begin{figure}[h]
  \centering
\includegraphics[scale=1.00]{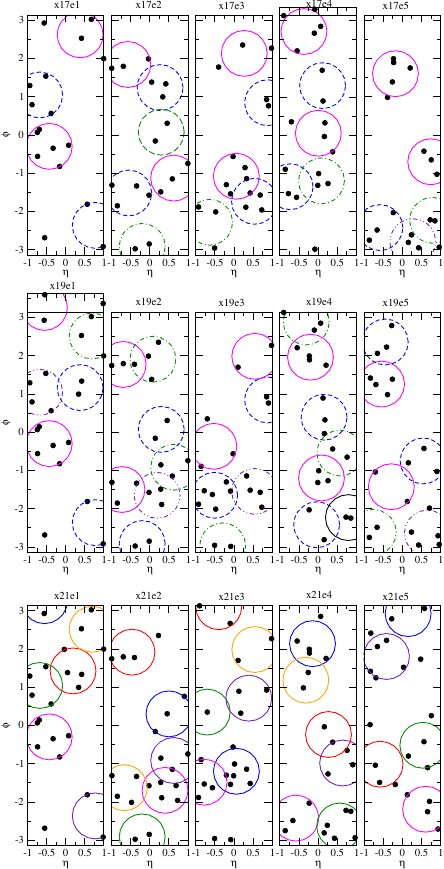}
\caption{Scatter plots in the $(\eta,\phi)$ plane for
  produced particles in events with multiplicities $M_{\rm random}$=17, 19, and 21
  generated an event generator with a uniform and independent
  production within  $|\eta|\le 1$ and $|\phi|$$\le$$ \pi$.
  Circular curves indicate the locations of the cluster circles with
  $R=0.6$.  }
  \label{fig12n}
\end{figure}
Thus, the upper limit of the number of clusters from the random
distribution Eq.\ (\ref{eq8}) increases quadratically as a function of
the multiplicity $M_{\rm random}$.  This upper limit can be quite
large for large $M_{\rm random}$.  For example, from
Eq.\ (\ref{eq16n}) one expects the upper limit of $K_{\rm
  random}(2,M_{\rm random})=$0.9 and 4.95 clusters for $M_{\rm
  random}$=5 and $M_{\rm random}$=11 respectively.  Thus, we would not
be surprised to find clusters even for randomly and independently
distributed particles as the multiplicity $M_{\rm random}$ increases
from 4 to 11.

In our numerical example, we generate particles randomly with the
uniform probability distribution of Eq. (\ref{eq8}) within $-\pi \le
\phi \le \pi$ and $-1 \le \eta \le 1$.  We label the events as
$xM_{\rm random}eI$ and show sample events with multiplicity from
$M_{\rm random}$=5 to $M_{\rm random}$=21 in Figs.\ \ref{fig10n}~
\ref{fig12n}, where we shall not distinguish the charges and the types
of particles.  We then use the minijet finding algorithm of Secs.
III and IV to locate cluster centers and circumscribe the cluster in
circles.

Figures\ \ref{fig10n} ~\ref{fig12n} show that as the multiplicity
increases, the number of clusters $K_{\rm random}$ also increases.  In
Figure \ \ref{fig9n}(a), we show that the number of clusters $K_{\rm
  random}$ appears to be nearly a linear function of the multiplicity,
similar to the relationship for events generated by PYTHIA 8.1.  The
number of clusters $K_{\rm random}$ estimated by Eq.\ (\ref{eq16n})
represents only an upper limit because a cluster with more than two
particles can be formed in high multiplicity events.  The number of
clusters increases only approximately linearly with multiplicity
$M_{\rm random}$, instead of the quadratic dependence of
Eq.\ (\ref{eq13n}), as shown in Fig.\ \ref{fig9n}(a).

Fig.\ \ref{fig9n}(a) shows the cluster numbers $K$ for events
generated by the random distribution, along with those generated by
PYTHIA8.1 within $|\eta|\le$ 1.  The number of clusters $K_{\rm
  random}$ for the random distribution appears to increase, likewise,
monotonically and approximately as a linear function of multiplicity
$M$=$M_{\rm random}$.  The relationship between $(M/K)_{\rm random}$
and $M$ is shown in Fig \ref{fig9n}(b).  The ratio
 $(M/K)_{\rm random}$ is 2.354 for M=5
and is 2.357 for M=8.  The ratio $K_{\rm PYTHIA}/K_{\rm random}$ is
close to unity.  It is 0.981 for M=5, and it is 0.996 for M=8, as shown
in Fig.\ \ref{fig9n}(c).

One way to study the clusters that are formed is by way of the
$(\Delta \eta=\eta_1-\eta_2,\Delta \phi=\phi_1-\phi_2)$ correlations
between clusters located at $(\eta_1,\phi_1)$ and $(\eta_2,\phi_2)$.
Figures.\ \ref{fig10n} ~ \ref{fig12n} for the random and uniformly
distributed particles, also exhibit azimuthal correlations for some of
the pairs, as cluster circles of similar types in these figures
indicate.  Thus, the clusters in the random distribution also exhibit
approximate azimuthal back-to-back correlations, as can be observed in
Figs.\ \ref{fig10n} and \ref{fig11n}.

We can estimate the number of azimuthally back-to-back correlated
  clusters $D'$ as a function of the number of clusters
$K_{\rm random}$.  We consider a pair of   clusters.  The probability
that the pair of   clusters can be considered back-to-back
correlated in azimuthal angles is
\begin{eqnarray}
P_{\rm random} = \frac{2R}{\Delta \phi_{\rm window} }=\frac{2R}{2\pi}.
\label{eq17n}
\end{eqnarray}
In an event with $K_{\rm random}$ number of   clusters,
the number of distinct   pairs is
\begin{eqnarray}
({\rm number ~of~ distinct~pairs}) = \frac{K_{\rm random}(K_{\rm random}-1)}{2}.~~~~~
\label{eq15n}
\end{eqnarray}
Therefore, in such an event with $K_{\rm random}$ number of clusters,
the (average) number of mini-dijet-ike pairs $D'(K_{\rm random})$ for
the random distribution is the product of Eqs.\ (\ref{eq14n}) and
(\ref{eq15n}),
\begin{eqnarray}
D'(K_{\rm random}) =\frac{K_{\rm random}(K_{\rm random}-1)}{2} \left( \frac{R}{\pi}\right ).~~~~~
\label{eq19n}
\end{eqnarray}
Thus, for $K_{\rm random}$=4, the number of mini-dijetlike pair is $D'(K_{\rm random}) $=1.15. 
This means that when $K_{\rm random}$ exceeds about four, the number of
mini-dijetlike pair of clusters $D'$$\sim$1 and back-to-back
correlated mini-dijet-like pair will begin to set in, as one can
observe from the number of mini-dijetlike clusters in events x11e3,
x13e2, and x15e1 with $K_{\rm random}\gtrsim 4$ in Fig.\ \ref{fig11n}.
 
Results in Figs. \ref{fig10n} ~\ref{fig12n} indicate that by
distributing particles densely within a small angular phase space,
clustering and azimuthal correlations occur also for randomly
distributed sources of particle.  Thus, clustering and azimuthal
correlation by themselves cannot be the only means of identifying
minijets and mini-dijets.  The identification of these clusters, as
such, arises from other independent supports for the dominance of the
hard-scattering model for minijet production of low-$p_T$ particles.

\section{ Correlations between particles and between clusters  \label{sec8}}

\subsection{Two-particle correlations}

The results in the previous sections indicate that while the PYTHIA
event generator yields clusters in the $(\eta,\phi)$ plane, such a
clustering is not uniquely a property of the dynamics of the particle
production processes in QCD as implemented in PYTHIA.  Clustering also
occurs with randomly generated data arising from many sources.  The
angular dimension $\pi R^2$ relative to the window dimensions $\Delta
\eta|_{\rm window}$ and $\Delta \phi|_{\rm window}$ gives rise to a
finite clustering probability even for random distributions, as
discussed in Eqs.\ (\ref{eq16n}), (\ref{eq17n}), and (\ref{eq19n}).
The kinematic cuts and the shape of the kinematic window also plays a
significant role.  With a particular geometry in defining the
acceptance window such that $\{\eta_1, \eta_2\} \in \Delta \eta_{\rm
  window}$, the phase space of a correlated particle-particle pair
will not always be distributed uniformly in the correlation
coordinates $\{\Delta \eta, \Delta \phi\}$.  As a consequence the
particle-particle correlations and the associated cluster-cluster
correlations will be distorted.  The assumed intrinsic property of the
clusters play another important role.  For example, if we set the
$p_T$ acceptance threshold of the particles to be higher and higher,
then the multiplicity number and the number of clusters will be 
lower, the greater the probability for a high multiplicity event to
   reveal itself much more readily as originating from a mini-jet as
compared to a random cluster.  Therefore, the meaning of a cluster is
defined by a given set of the attributes of the delimiting
constraints.

Given a set of these constraints, we would like to examine the
particle-particle and cluster-cluster correlations for the purpose of
extracting information on the dynamics that distinguishes the
PYTHIA 8.1 results from random results.  The particle-particle and
cluster-cluster correlations are also called two-particle and
two-cluster correlations respectively.  After using PYTHIA 8.1 or the
random distribution to simulate $pp$ collisions, we collect the
kinematic data of the particles in each event.  We pick all
combinations of particle pairs (or cluster pairs) that are in the same
events to calculate the $\Delta\phi$ and $\Delta\eta$ between any two
particles or clusters. Then we fill the 2D-histogram with $\Delta
\phi$ and $\Delta \eta$ to get the particle-particle or
cluster-cluster correlation function.

\begin{figure}[h]
\includegraphics[width=7.5cm,height=11.6cm]{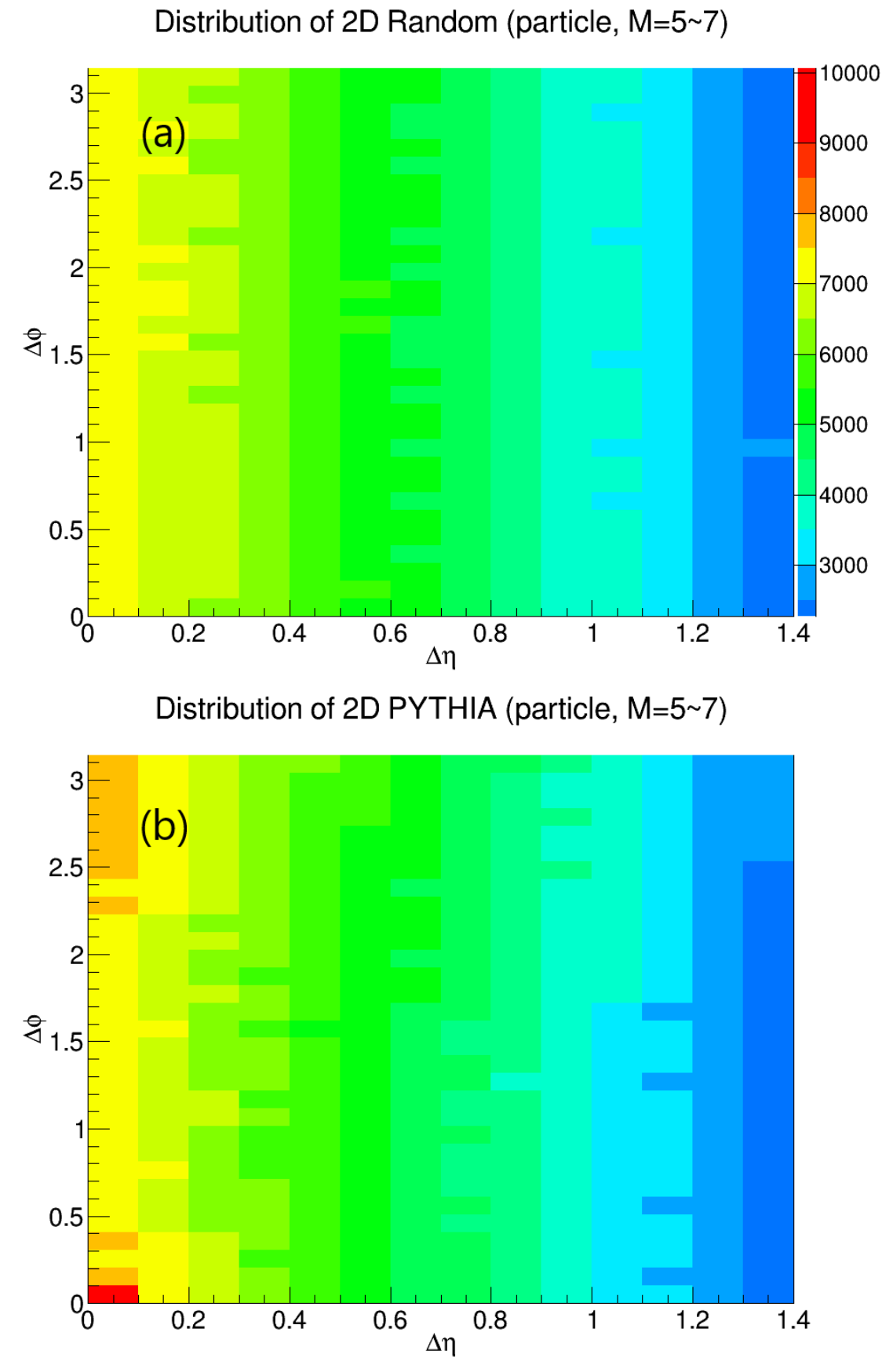}
\caption{The unnormalized 2D two-particle correlation
  distribution $dN/(d\Delta \phi d\Delta \eta)\big |_M$ is produced
  by the events with multiplicities $M$ =5, 6, and 7.  Fig. (a) is the
  cluster-cluster correlation for the random events and Fig. (b) is
  the cluster-cluster correlation for PYTHIA events. The $\Delta \phi$
  and $\Delta \eta$ are positive as they are only differences in the
  azimuthal and pseudorapidity coordinates.}
\label{fig13n}
\end{figure}

In Fig.\ref{fig13n}, we show the unnormalized 2D two-particle
correlation distribution $dN/d\Delta \eta d \Delta \phi$ as a function
of the correlation separation $\Delta \eta$ and $\Delta \phi$ within
the window of $-1\le \eta\le 1$ and $0\le \phi\le \pi$ for $M$=5-7.
There is the symmetry of the distribution with respect to a change of
the sign of $ \Delta \phi$ or $ \Delta \eta$.  It suffices to display
the distributions only in the region of positive $ \Delta \phi$ and $
\Delta \eta$.  The color plot of the event number in each bin is
red for large number counts and blue for fewer number of counts.

The effect of the phase space limitations shows up clearly in the
correlation function for the case of the random distribution in
Fig.\ \ref{fig13n}(a).  Within the acceptance window, the generated
correlation function $dN/d{\Delta \phi d\Delta \eta}$ along the $\eta$
axis is large at $\Delta \eta=0$, and it falls down linearly as $\Delta
\eta$ increases in the well-known form of a triangular distribution.
The generated correlated distribution is nearly uniform in the $\Delta
\phi$ direction with minor fluctuations.

Figure \ref{fig13n}(b) gives the two-particle correlation obtained
with the event generator PYTHIA 8.1, for $M$=5-7, which corresponds
approximately to the average multiplicity $\langle M \rangle$ =6.94.
It represents essentially the theoretical particle-particle
correlation function for the case of minimum-biased measurements.  One
notes that the limited phase space of the measurement window, likewise,
distorts the distribution to follow roughly the triangular shapes as a
function of $\Delta \eta$, with an approximately uniform distribution
in $\Delta \phi$.  However, upon careful examination, there are
finer differences in the region of $(\Delta \phi, \Delta \eta)\sim 0$
and $\Delta \phi \sim \pi$.  Along the $\Delta \phi$ axis,
Fig.\ref{fig13n}(b) shows two peaks at $\Delta \phi \sim$ 0 and
$\Delta \phi \sim \pi$.

\begin{figure}[h]
\includegraphics[width=7.5cm,height=15cm]{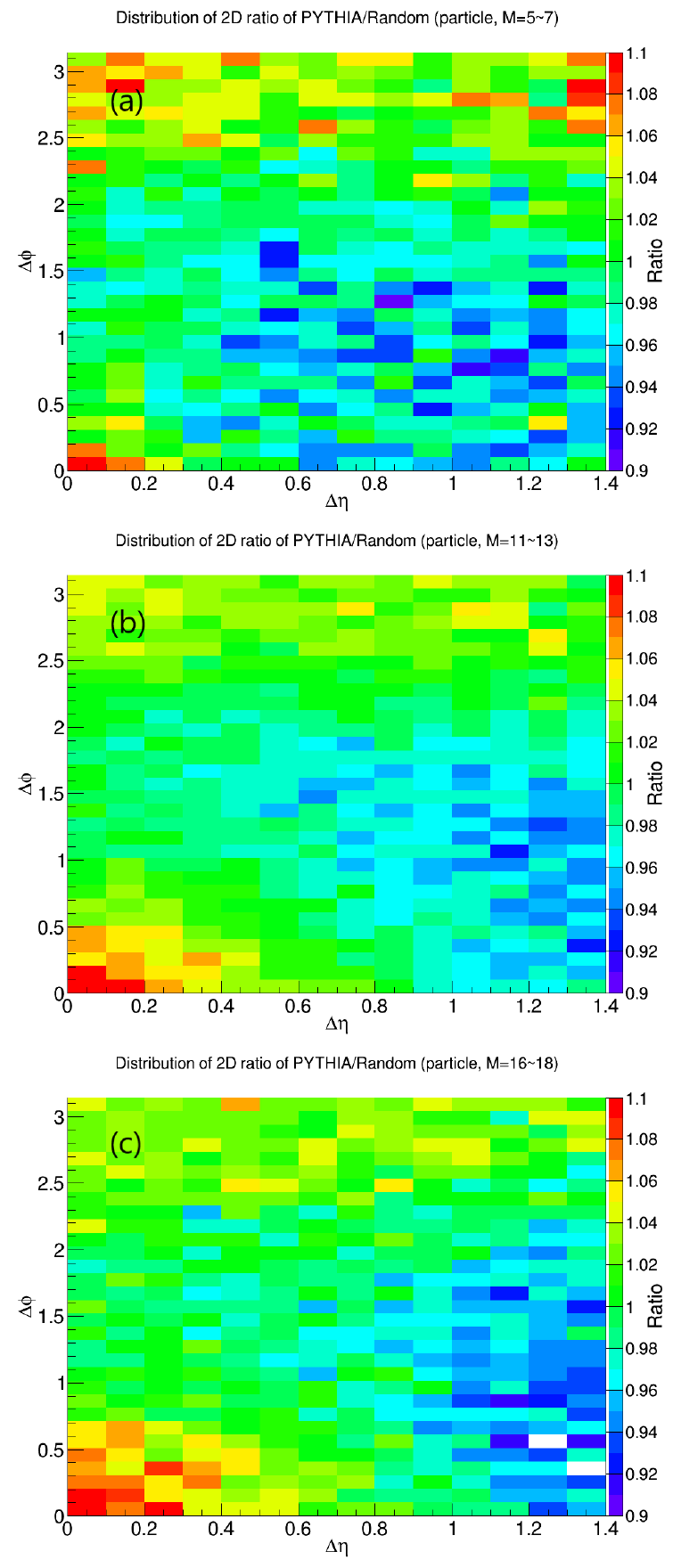}
\caption{2D normalized two-particles correlation
  function $(dN/d\Delta \phi d\Delta \eta)_{(PYTHIA)}/(dN/d\Delta \phi
  d\Delta \eta)_{(Random)}$. The figures are produced by events with
  $M$ =11, 12, 13. The horizontal coordinate is $\Delta \eta$, and the
  vertical coordinate is $\Delta \phi$.}
\label{fig18n}
\end{figure}

The difference between the PYHTIA distribution and the random
distribution shows up in sharper focus upon taking the ratio
$dN/d\Delta \eta d \Delta \phi|_{\rm PYTHIA}/dN/d\Delta \eta d \Delta
\phi|_{\rm random}$ at each $(\Delta \eta, \Delta \phi)$ point as
shown in Fig.\ \ref{fig18n}(a).  We shall call such ratio,
$dN/d\Delta \eta d \Delta \phi|_{\rm PYTHIA}/dN/d\Delta \eta d \Delta
\phi|_{\rm random}$, the normalized correlation function.  It is
normalized with respect to the constraints of the measurement as
represented by a random distribution within the measurement window.
In the comparison with experimental data, such normalizations are often
carried out by event-mixing data.

The normalized 2D particle-particle correlation obtained with the
PYTHIA 8.1 generator exhibits the gross feature of the well-known shape
with a near-side peak at $(\Delta \eta, \Delta \phi) \sim 0$ arising
from the near-side jet and an away-side ridge at $\Delta \phi \sim
\pi$ along the direction of approximately constant $\Delta \eta$ by
momentum conservation in a parton-parton collision
[55-63,80,92,93].  Beyond
these two regions, there is a region of low correlations with $\Delta
\eta > 0.6$ and $0 < \Delta \phi < 1.5$.

We can carry out similar analysis for other values of the multiplicity
number $M$.  We find out that particle-particle correlations change as
the multiplicity $M$ increases.  In Fig.\ \ref{fig18n}(b), we show the
particle-particle correlation for $M=11-13$ for which one finds that
as the multiplicity increases the near-side jet gains in strength and
angular size, and the away side correlation becomes weaker because
momentum conservation is weakened by a larger multiplicity.

In Fig.\ \ref{fig18n}(c), we show the particle-particle correlation
for $M=17-18$, for which one finds that for such higher multiplicity
events the near-side jet gains even greater in strength and angular
size, and similar to the case of $M=11-13$ the away side ridge
distribution cannot be distinguished.

\subsection{Two-cluster Correlations}

While the particle-particle correlations comprise a part of the
standard tools in the analysis of experimental data, it is of interest
to develop cluster-cluster correlations as another useful tool in the
study of the dynamics of the particle production process.
Accordingly, for a set of delimiting cut-off attributes for accepting
a particle and a cluster, we apply the cluster-searching algorithm to
locate the clusters and their centers.  Because each cluster has at
least two particles, we start to study events with at least five particles
so that there are sufficient number of particles and clusters to
examine cluster-cluster correlations.  With the knowledge of the
cluster centers, we pick all combinations of particle pairs that are
in the same event to calculate the $\Delta\phi$ and $\Delta\eta$
between the centers of any two clusters and obtain the cluster-cluster
correlation.

\begin{figure}[h]
\includegraphics[width=7.5cm,height=11.6cm]{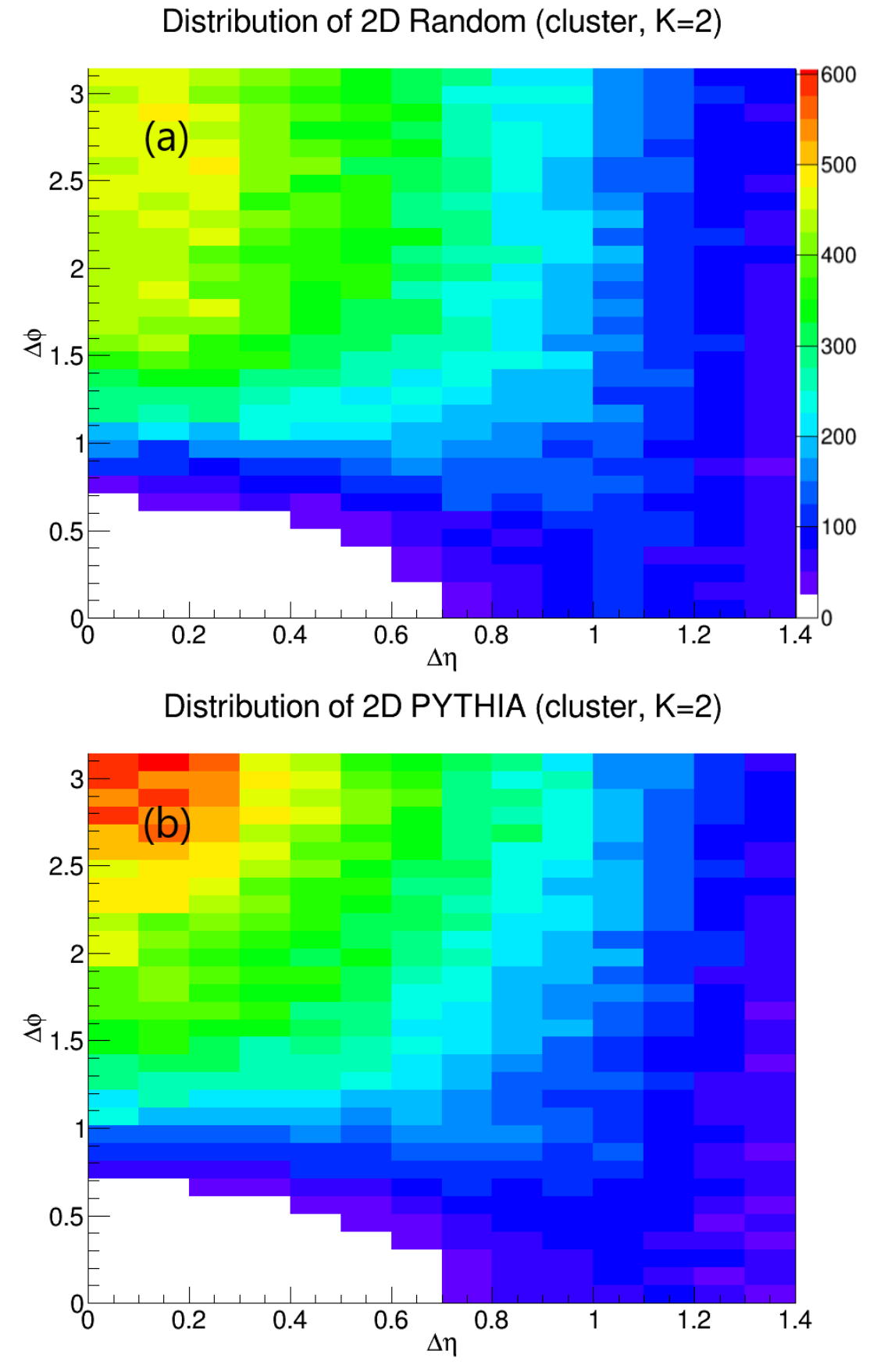}
\caption{The unnormalized 2D two-cluster correlation
  distribution $dN/(d\Delta \phi d\Delta \eta)$ is produced by
  events with 2 clusters.   Figure (a) is the two-cluster $\Delta \eta$-$\Delta \phi$
  correlation function for events from the random distribution, and
Fig. (b) is the two-cluster $\Delta
  \eta$-$\Delta \phi$ correlation function for events generated by
  PYTHIA8.1.  The
  $\Delta \phi$ and $\Delta \eta$ are positive as they are only
  differences of the azimuthal and pseudorapidity coordinates of the
  pair of clusters.}
\label{fig15n}
\end{figure}

Figure \ref{fig15n}(a) is the 2D two-clusters correlation distribution of
the $\Delta \phi$ and $\Delta \eta$ with K = 2 for the random
distribution.  This case of $K=2$ corresponds approximately to the
case of $M$=5-7 as shown in Fig.\ \ref{fig13n}.  We can understand the
gross features of the cluster-cluster correlation in
Fig.\ref{fig15n}(b) in the following way.  In the region of $\Delta
\eta\sim 1.4$, the correlations for the random case is small, arising
from the phase-space limitation of the triangular distribution of the
particle-particle correlations as shown in \ref{fig15n}(a).  As
$\Delta \eta$ decreases below the cluster radius $R$, those particles
falling within the domain of the first cluster within a radius $R$
will become part of the other cluster, and thus, the probability of
another cluster in the $(\Delta \eta,\Delta \phi) < R$ region of the
first cluster is essentially zero when $K=2$ as indicated by a void in
the $(\Delta \eta,\Delta \phi) \sim 0 $ region.  For the region at
($\Delta \eta\sim 0$,$\Delta \phi\sim \pi$), there is a natural
enhancement of the correlation because the relatively large value of
the correlation function at $\Delta \eta \sim$0 that enhances the
formation of clusters at $\Delta \eta \sim$0 and $\Delta \phi\sim\pi$.

\begin{figure}[h]
\centering
\includegraphics[width=8cm,height=15cm]{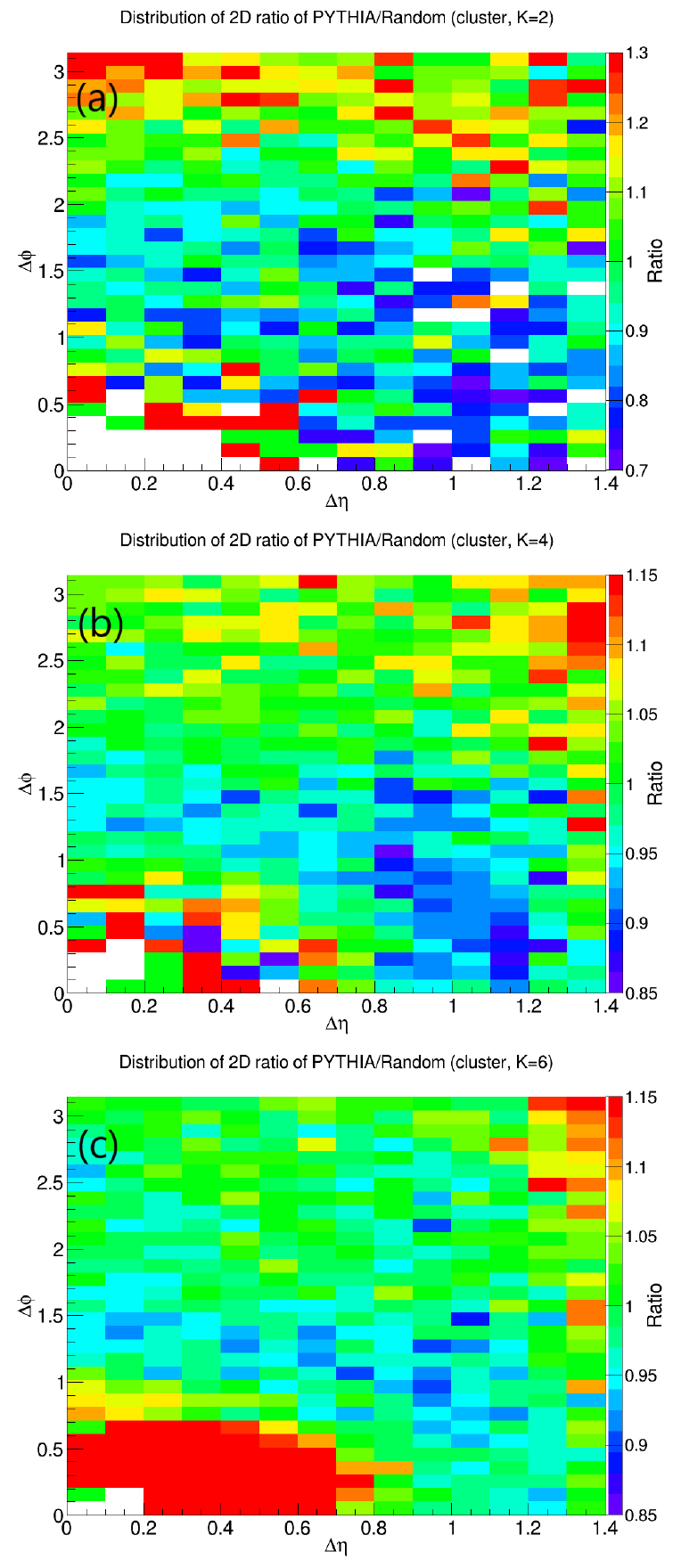}
\caption{Normalized 2D two-clusters correlation with function $(dN/d\Delta \phi d\Delta \eta)_{(PYTHIA)}/(dN/d\Delta
  \phi d\Delta \eta)_{(Random)}$. The figures are produced by events
  with 2, 4 and 6 clusters. The horizontal coordinate is the ratio of
  $\Delta \eta$, and the vertical coordinate is the ratio of $\Delta
  \phi$.}
\label{fig16n}
\end{figure}

Figure \ref{fig15n}(b) gives the 2D two-clusters correlation distribution
of the $\Delta \phi$ and $\Delta \eta$ with $K = 2$ obtained with
particles generated from by PYTHIA 8.1.  This case of $K=2$ corresponds
to the case of $M$=5-7 as shown in Fig.\ \ref{fig13n}(b) and
represents approximately the minimum-biased case.  The cluster-cluster
correlation for the PYTHIA 8.1 case retains some of the gross features
as presented from the random case, arising from the phase-space
limitation of the measurement windows.  There is, however, finer
differences arising from the dynamics of the PYTHIA 8.1 particle
generating processes.  There appears to be an enhancement of the
distribution at $\Delta \phi\sim \pi$, reflecting the occurrence of
the back-to-back nature of the fragmentation processes as implemented
in the PYTHIA 8.1 program.  The fine details show up more clearly upon
taking the ratio $dN/d\Delta \eta d \Delta \phi |_{\rm PYTHIA}/
dN/d\Delta \eta d \Delta \phi |_{\rm random}$ over the $(\Delta
\eta,\Delta \phi)$ plane.  Such a ratio will be called the normalized
cluster-cluster correlation function for PYTHIA 8.1 as show in
Fig. \ref{fig16n}(a).

The 2D normalized cluster-cluster correlation functions for $K=2, 4,$
and 6 in Fig. \ref{fig16n} exhibits the dynamics and its variation of
the PYTHIA calculations as the cluster number changes.  For $K=2$, the
correlation shows a near-side jet in the region of $(\Delta
\eta,\Delta \phi)$ $\sim$ 0.  This near-side jet grows in strength as
the cluster number $K$ increases.  On the other hand, at $\Delta \phi
\sim \pi$, there appears to be away-side ridge along the $\eta$ direction
for $K=2$.  As $K$ increases, the ridge feature is modified to become
a peak at $(\Delta\eta$$\sim$1.4, $\Delta\phi $$\sim $$\pi)$,
separated from the peak at $(\Delta \eta, \Delta \phi )$$\sim$ 0.

It is instructive to compare the particle-particle corrections in
Fig.\ \ref{fig18n} with their corresponding cluster-cluster
correlations in Fig.\ \ref{fig16n}.  For the case of $M$=5-7 that
corresponds closely with the case of $K=2$, one observes in both
particle-particle and cluster-cluster correlation the occurrence of
the near-side jet and the away-side ridge.  The cluster-cluster
correlations yields an amplified amplitude of about 30 percent,
whereas the particle-particle correlation yields an amplitude of only
about 10 percent.  Thus, in comparison with the particle-particle
correlation, the cluster-cluster correlation amplifies the amplitude
for the away-side ridge to a greater degree.  For the case of
$M=10-13$ in comparison with $K=4$, the correlation at
$(\Delta\eta,\Delta \phi) \sim 0$ is enhanced while the
particle-particle correlation does not exhibit a large enhanced
amplitude at that location; the cluster-cluster correlation exhibits
an enhanced amplitude at $\Delta \eta \sim 1.4$ and $\Delta \phi\sim
\pi$. For the case of $M=17-19$, or the corresponding case of $K=6$,
the correlation at $(\Delta\eta,\Delta \phi)\sim 0$ is even more
enhanced.  Furthermore, the correlation function appears to be greatly
enhanced at $(\Delta \eta\sim 1.4, \Delta \phi\sim \pi)$ indicating a
regularity of the dynamics at a certain $\Delta \eta\sim 1.4$ in the
away-side angles.

For completeness, there are additional pieces of information one can
gain from the results of $dN/d\phi$ and $dN/d\eta$ distributions which
we shall present in the Appendix.

\section{Conclusions and Discussions}

The parton-parton hard scattering is an important process in
high-energy nucleon-nucleon collisions.  Although originally conceived
to involve only the production of high-$p_T$ particles, it has been
suggested that the dominance of the hard-scattering process may extend
to the low-$p_T$ region with the production of minijets and
mini-dijets, as the collision energy increases.
  
As a first attempt to identify minijets, we develop an algorithm to
search for clusters using the k-means clustering method, supplemented
with a k-number (cluster-number) selection principle.  The method
adopts a scheme of random initialization of the initial centers,
minimizing the potential function $\Phi(K)$ for a fixed $K$, and
looking for the $K$ number of clusters that leaves the fewest number
of particles $\Omega$ outside the cluster circles.  The method is
stable, fast, and yields clusters and their associated particles.

Using such a method, we have located clusters in the $(\eta,\phi)$
plane on an event-by-event basis, using events generated by PYTHIA 8.1,
which contains the dynamics of multiple parton interactions.  To a
cluster identified by the procedures, one often finds an associated
cluster located at approximately $|\Delta \phi_{\rm jet-jet}| \sim
\pi\pm R$.  Their azimuthal angular correlation suggests that they may
be identified as the two partners of a mini-dijetlike pair.  We find
that clusters of low-$p_T$ hadrons are common occurrences for
PYTHIA 8.1 events with high multiplicities.  The number of multiple
clusters increases approximately linearly with increasing multiplicity
$M$.

It must be pointed out, however, that clustering and azimuthal
correlations alone cannot be the only means to identify minijets and
mini-dijets.  A randomly distributed set of particles in large
multiplicities also exhibit clustering properties similar to those
from the PYTHIA 8.1 program with minijets.  The ability to distinguish
the dynamics of the particle production processes will require the
measurement of the particle-particle and cluster-cluster correlations.

We have examined the particle-particle and cluster-cluster
correlations obtained with particles generated from PYTHIA and
compared these correlations with those from the random distribution.
We need to normalize these correlations properly by dividing the
correlation function obtained in PYTHIA by the correlation function
obtained by a random distribution.  We find that the normalized
correlation function from PYTHIA has features that distinguish
themselves from those of a random distribution.  In this regard, the
quantitative assessment of the dominance of the relativistic
hard-scattering process in the low-$p_T$ region needs to be
independently established in order to identify the clusters as
physical minijets.  The success of such an identification will
provide a tool to investigate minijet and mini-dijet properties, for
which not much detailed information has been collected.  Furthermore,
quantitative predictions based on first principles of perturbative QCD
for the low-$p_T$ region is difficult because the multiple collision
probability involves higher-order corrections beyond the leading order
\cite{Kot17}.

From our investigations, one may also wish to develop strategies to
apply the proposed algorithm to examine experimental data at various
energies and examine information on the production cross sections and
the phase-space distribution of these objects, for comparison with the
theory of multiple minijet production as a function of the collision
energies.  In this regard, we should note that the higher the $pp$
collision energy, the greater the probability is of the dominance of
the hard-scattering process for the production of low-$p_T$ particles,
and the greater the probability will be of the clusters to be physical
minijets.
 
We have introduce a general method only as a first step towards
our eventual goal of locating minijets.  In future practical
analysis in the search for minijets in experimental data, it may
    be reasonable to include a supplementary requirement that a
    minijet must contain at least a single particle with a $p_T$
    greater than a certain threshold value $p_0$ (say 1 GeV/c).  Such
    a supplementary condition will serve the good purpose of fixing
    the minijet cluster number $K$ to facilitate the searching
    algorithm (in place of the present principle of the least number
    of outside points).  It will reduce the number of clusters so that
    the effects of the finite size of the experimental windows is
    reduced.  It will also bring us closer to the goal of examining
    minijets, multiple minijets, and their correlations.  Future
    investigation along such directions, in conjunction with a properly
    modified k-means clustering algorithm, will be of great interest.

\vspace*{0.50cm}

\centerline{\bf Acknowledgments}
\vspace*{0.5cm} The authors would like to thank Professors  Zhenyu Ye and
Soren Sorensen for helpful discussions.  The research was supported in
part by the Division of Nuclear Physics, U.S. Department of Energy
under Contract DE-AC05-00OR22725 with UT-Battelle and Contract
No.DE-FG02-88ER40424 with UCLA.

\appendix

\section{ The Elbow  Method of     Cluster Number Selection}

\begin{figure}[h]
  \centering
\includegraphics[scale=0.45]{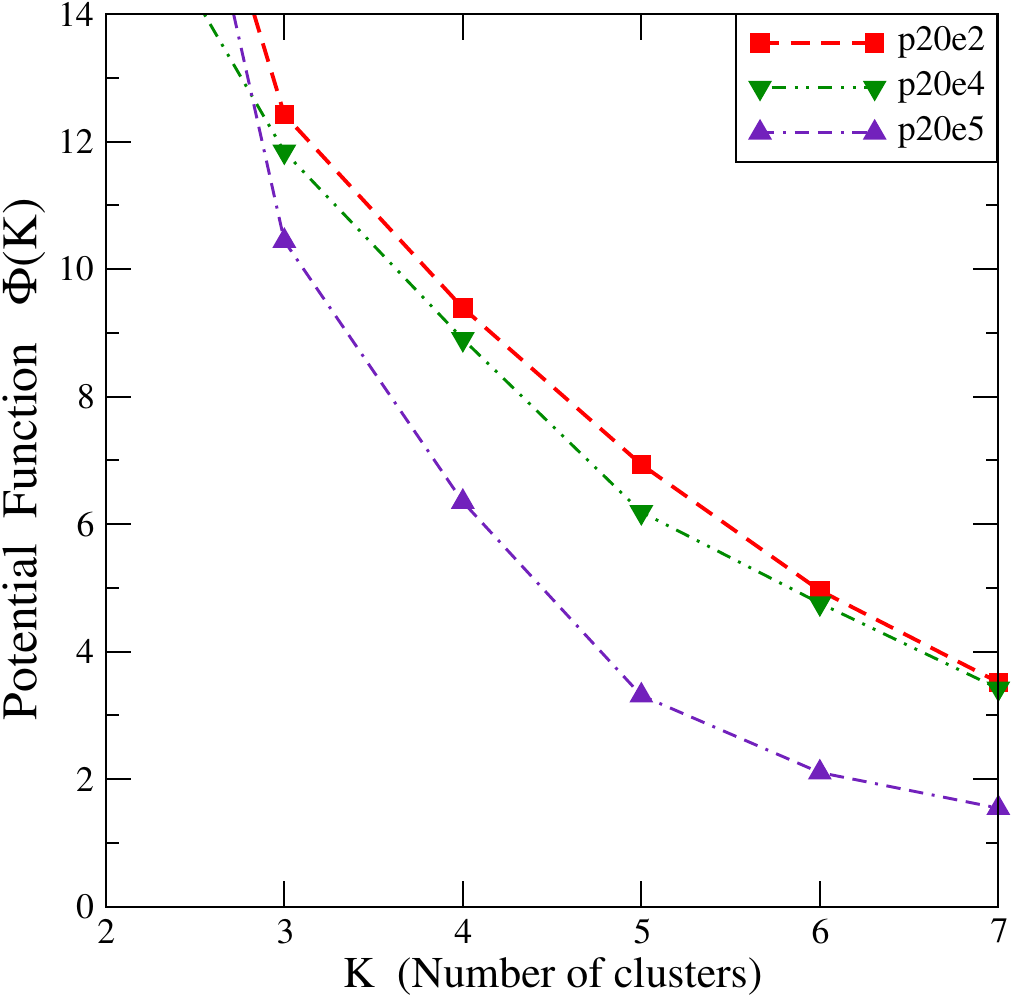}
\vspace*{-0.4cm}
\caption{Potential function $\Phi(K)$, the sum of
  square distances between the subset data points and their
  corresponding cluster centers, as a function of the number of
  clusters $K$, for minimum-bias events with multiplicity $M=$ 20
  generated by PYTHIA 8.1.  }
\label{fig3new}
\end{figure}

There is another method to select the cluster number $K$ by studying
the $K$-dependence of the potential function $\Phi(K)$.  For a given
$K$ value, after the minimization of the potential function $\Phi(K)$
with respect to the random initialization of the cluster centers and
the variations of the cluster center positions, the quantity $\Phi(K)$
of Eq.\ (\ref{eq5}) is then evaluated.  The potential function
$\Phi(K)$ is on the whole a decreasing function of increasing $K$
(Fig.\ \ref{fig3new}), as it reaches the limiting value of zero when
the number of clusters $K$ is the same as the number of data points
$M$.  An inefficient and  a slowly decreasing function  of $\Phi(K)$ occurs, if a
cluster is subdivided into smaller subclusters with a subsequently
smaller change of the $\Phi(K)$ slope.  On the other hand, a large and
abrupt change of $\Phi(K)$ as a function of $K$ signifies a
significant change of the structure of the clustering configuration
and may be the location of the appropriate cluster number.  Hence, it
has been suggested that the proper cluster number $K$ occurs at the
kink (or elbow) of the curve of $\Phi(K)$ as a function of $K$ or at
the location of an abrupt change of the slope of $\Phi(K)$
\cite{Ng17,Tho53}.

We calculate the potential function $\Phi(K)$ as a function of the
cluster number $K$ for events with $M=$ 20 as shown in
Fig.\ \ref{fig4n}.  For event p20e2 shown in Fig.\ \ref{fig4n}, a kink
of $P(K)$ occurs at $K$=3 and a very weak kink also appears to occur
at $K$=6.  The determination of the location of the kink is not
without ambiguity.  The elbow method would suggest the cluster number
of $K$=3 or 6 but as we observed in Fig.\ \ref{fig4n}, the proper
cluster number as determined from the principle of fewest outside
points is $K=$7.  For event p20e4, kinks of $\Phi(K)$ occur at $K$= 3
and 5, but the appropriate cluster number as determined from the
principle of fewest outside points is 7.  For p20e5, the potential
function shows a sharp kink at $K$=3 and weaker kinks at 5 and 6,
whereas the method of the principle of fewest outside points gives
$K$=6.  The method of the sharpest kink has the difficulty of
recognizing the location of the kink, as many changes of slopes occur
at different locations.  If one takes the method to be given by the
location with the greatest change of the magnitude of the slope, it
would give $K$ numbers, which differ from the k-number selection
principle of the fewest number of outside points.

We conclude that in the elbow method the determination on the location
of the kink is ambiguous and there is no obvious method to resolve
the ambiguities.  The principle of the fewest outside points should
be the proper criterion for the selection of the proper cluster number
$K$ as it is based on the physical property of the clustering of a
minijet.

\section{Two-particle and two-cluster $dN/d\Delta\eta$ and $dN/d\Delta\phi$}

The results in Sec. \ref{sec8} provide a wealth of information on the particle-particle and cluster-cluster correlation functions ${dN}/{d\Delta\eta d\Delta\phi}$ in the $(\Delta\eta,\Delta\phi)$ plane. The shape of the correlation function landscape contains many interesting features.
In many measurements, it may be useful also to collect information on the ``marginal distributions" $dN/d\Delta \eta$ and $dN/d\Delta \phi$ by integrating the two-dimensional distribution $dN/d\Delta \eta d \Delta \phi$ over $\Delta \eta$ or $\Delta \phi$ directions as
\begin{eqnarray}
\frac{dN}{d\Delta \eta}=\sum_{\Delta \phi}^{} \frac{dN}{d\Delta \eta d\Delta \phi} d{\Delta \phi},
\nonumber\\
\text{and}~~~ \frac{dN}{d\Delta \phi}=\sum_{\Delta \eta}^{} \frac{dN}{d\Delta \eta d\Delta \phi} d{\Delta \eta}.
\nonumber
\end{eqnarray}

\begin{figure}[h]
\centering
\includegraphics[width=7.5cm,height=9cm]{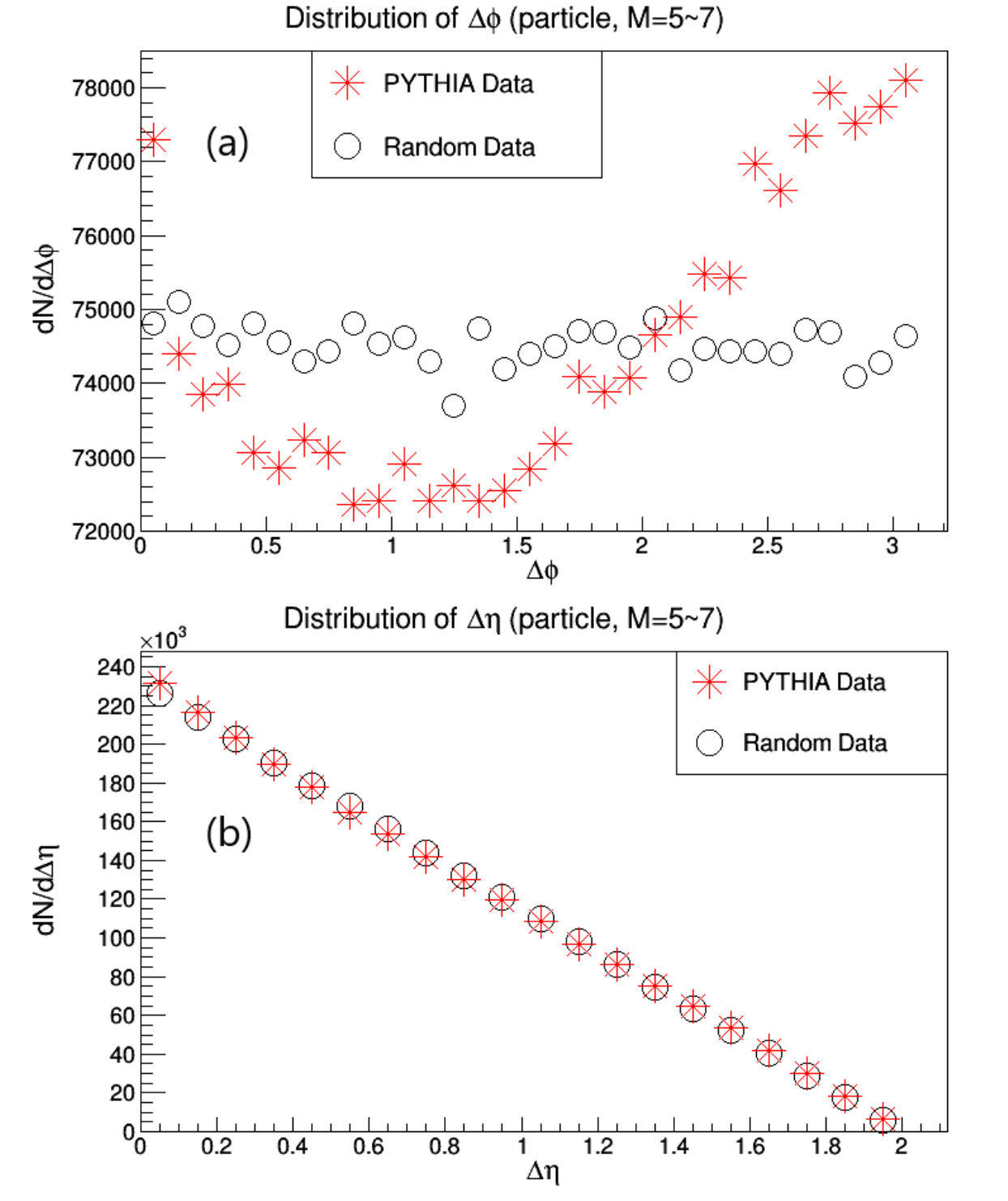}
\caption{Un-normalized two-particle differential correlation distribution 
 (a)  $dN/d\Delta\phi$ and  (b) $dN/d\Delta\eta$ ,
 for events with multiplicity $M = 5 \sim 7$.  The red inverted triangle points are  for  events generated by PYTHIA 8.1 and  the black circular points are from events generated from a random distribution. 
}
\label{fig21n}
\end{figure}
These marginal distributions are not as informative as the full 
two-dimensional $dN/d\Delta \eta d \Delta \phi$ distribution
because  many important features may become obscured
when the two dimensional distribution has been integrated. They provide partial information on the 
particle or cluster distribution projected in certain directions.
Recognizing that these marginal distributions are only part of the full distribution,
we shall show here the  marginal distributions  for various values of particle numbers $M$ and cluster number $K$, for completeness.

\vspace*{0.5cm}
\centerline{\bf B.1 Two-Particle  $dN/d\Delta\eta$ and $dN/d\Delta\phi$}
\vspace*{0.5cm}

\begin{figure}[h]
\centering
\includegraphics[width=7.5cm,height=17.5cm]{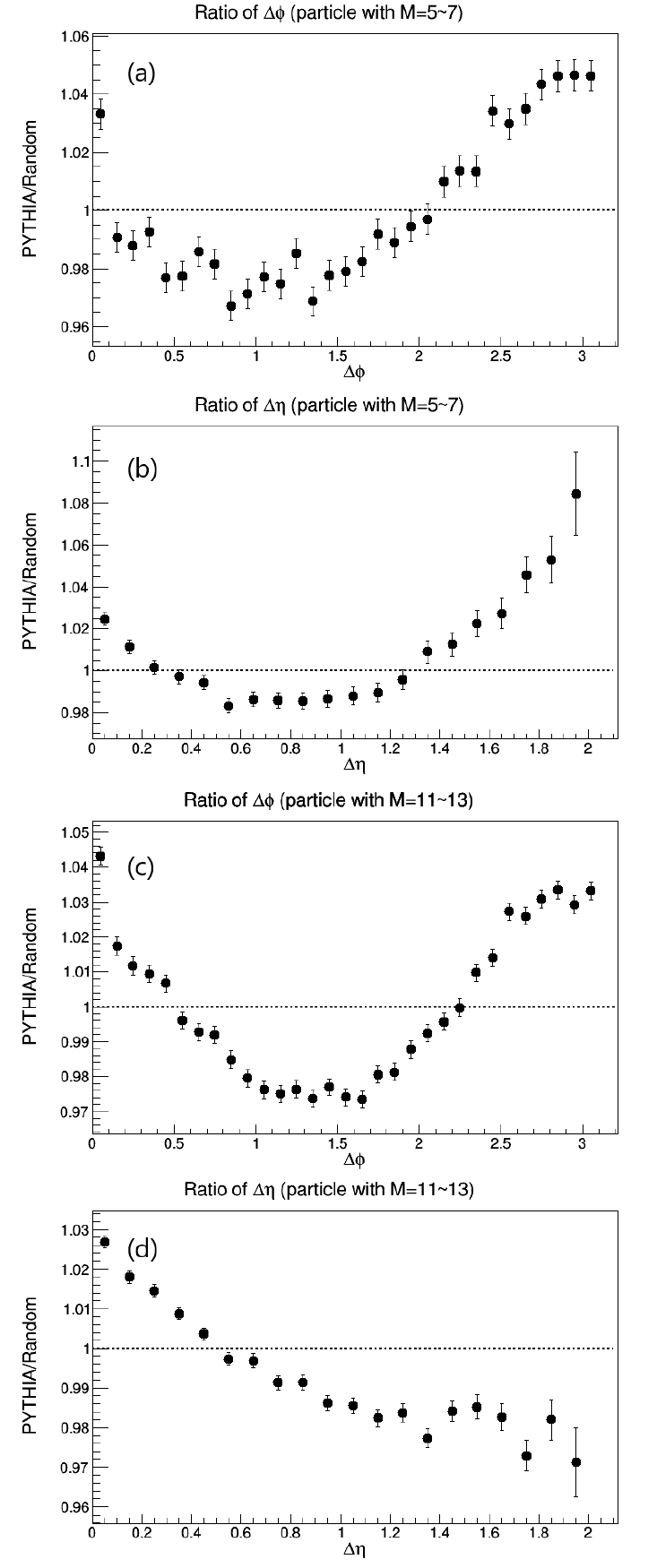}
\caption{Normalized two-particles correlation function  with $M= 5 ~{\rm to~}7$ and $= 11~{\rm to}~13$. Figures (a) and (c) are $\Delta\phi$ with function $(dN/d\Delta\phi)_{\rm PYTHIA}/(dN/d\Delta\phi)_{\rm random}$ and Figs. (b) and (d) are that of $\Delta\eta$ with function $(dN/d\Delta\eta)_{\rm PYTHIA}/(dN/d\Delta\eta)_{\rm random}$.}
\label{fig22n}
\end{figure}

Fig.\ \ref{fig21n}  gives  the un-normalized two-particle correlation distribution with $M \sim 5-7$.  The quantity $dN/d\Delta \phi$ in Fig.\ \ref{fig21n}(a) shows a nearly flat  distribution for the random distribution but a back-to-back correlation for the PYHTIA 8.1 calculations.  The quantity $dN/d\Delta \eta$ in Fig.\ \ref{fig21n}(b) shows a triangular distribution because of the $\eta$ window.  The difference between the distributions from PYTHIA 8.1 and from the random distribution is small.

To illustrate the finer differences between the distributions,  
we defined normalized distributions as
\begin{eqnarray}
{\rm normalized}~ \frac{dN}{d\Delta \eta}=\frac{ \frac{ dN}{d\Delta \eta|_{\rm PYTHIA}}}
{\frac{ dN}{d\Delta \eta |_{\rm random}}}
\end{eqnarray}
and 
\begin{eqnarray}
{\rm normalized}~ \frac{dN}{d\Delta \phi}=\frac{ \frac{ dN}{d\Delta \phi|_{\rm PYTHIA}}}
{\frac{ dN}{d\Delta \phi |_{\rm random}}}
\end{eqnarray}

 We display the normalized distribution for $\Delta\phi$ and $\Delta\eta$
for $M\sim 5-7$ in 
Fig.\ref{fig22n}(a) and (b) 
and for  $M\sim 11-13$ in Fig.\ref{fig22n}(c) and (d).
The
minijet component shows up  as a peak  at $\Delta \phi \sim 0$ in $dN/d\Delta \phi$ in 
Fig.\ \ref{fig22n}(a) and (c).
The 
away-side back-to-back correlation appears  as a rising peak of
$dN/d\Delta \phi$
at $\Delta \phi\sim \pi$  in Fig.\ \ref{fig22n}(a) and (c).   
The  near-side jet shows up as peak at $\eta\sim 0$ in $dN/d\Delta \eta$ and shows up at both $M\sim 5-7$ 
in Fig.\ \ref{fig22n}(b) and (d).   The $dN/d\eta$ at large $\Delta \eta$ 
gives a peak at large $\eta$ for $M\sim 5-7$ but the distribution decreases for large values of $M$ in Fig.\ \ref{fig22n}(d).  The away-side peak has a smaller magnitude as $K$ increases, expected by the dilution effect of momentum conservation.

\newpage

\vspace*{0.5cm}
\centerline{\bf B.2 Two-Cluster  $dN/d\Delta\eta$ and $dN/d\Delta\phi$}
\vspace*{0.5cm}

\begin{figure}[h]
\centering
\includegraphics[width=7.5cm,height=9cm]{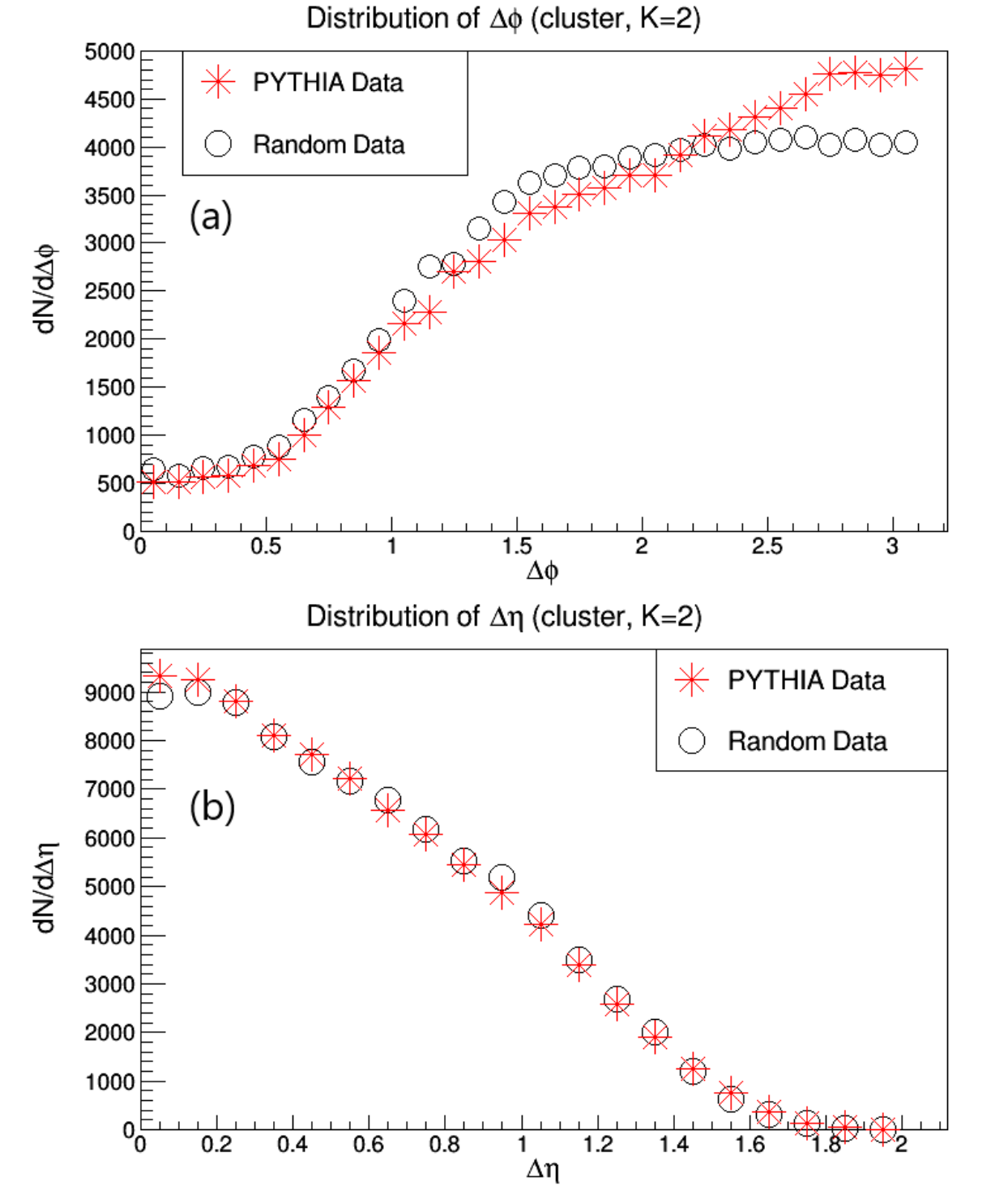}
\caption{
Un-normalized two-cluster differential correlation distribution 
 (a)  $dN/d\Delta\phi$ and  (b) $dN/d\Delta\eta$ ,
 for events with multiplicity $M = 5 \sim 7$.  The red star ``$*$"  points are  for  events generated by PYTHIA 8.1 and  the (black) circular points are from events generated from a random distribution. 
}
\label{fig23n}
\end{figure}

Figure \ref{fig23n}(a) and (b) give  the un-normalized two-clusters correlation distribution  $dN/d\Delta \phi$ and $dN/d\Delta \eta$ respectively,  for $K = 2$.
The $dN/d\Delta \phi$ distribution is nearly flat for $\Delta\phi\sim \pi$, but suppressed near the region near $\Delta \phi\sim 0$ because particles near $(\Delta\eta,\Delta \phi)\sim 0$ becomes part of the first cluster and there cannot be a second cluster nearby.    Similar behavior occurs for the distribution from PYTHIA events.   The two distributions have slightly different behaviors at $\Delta \phi \sim \pi$, arising from the occurrence of back-to-back correlations in dijet events.
The distribution in 
Fig.\ref{fig23n}(b) for $dN/d\Delta \eta$ is similar to the triangular distribution 
in Fig.\ \ref{fig21n}(b), except that the distribution decreases faster to zero at a smaller value of $\Delta \eta \sim1.6$ instead of $\Delta \eta \sim2.0$ 
in Fig.\ \ref{fig21n}(b).

\begin{figure*}[ht]
\centering
\includegraphics[width=\textwidth,height=8cm]{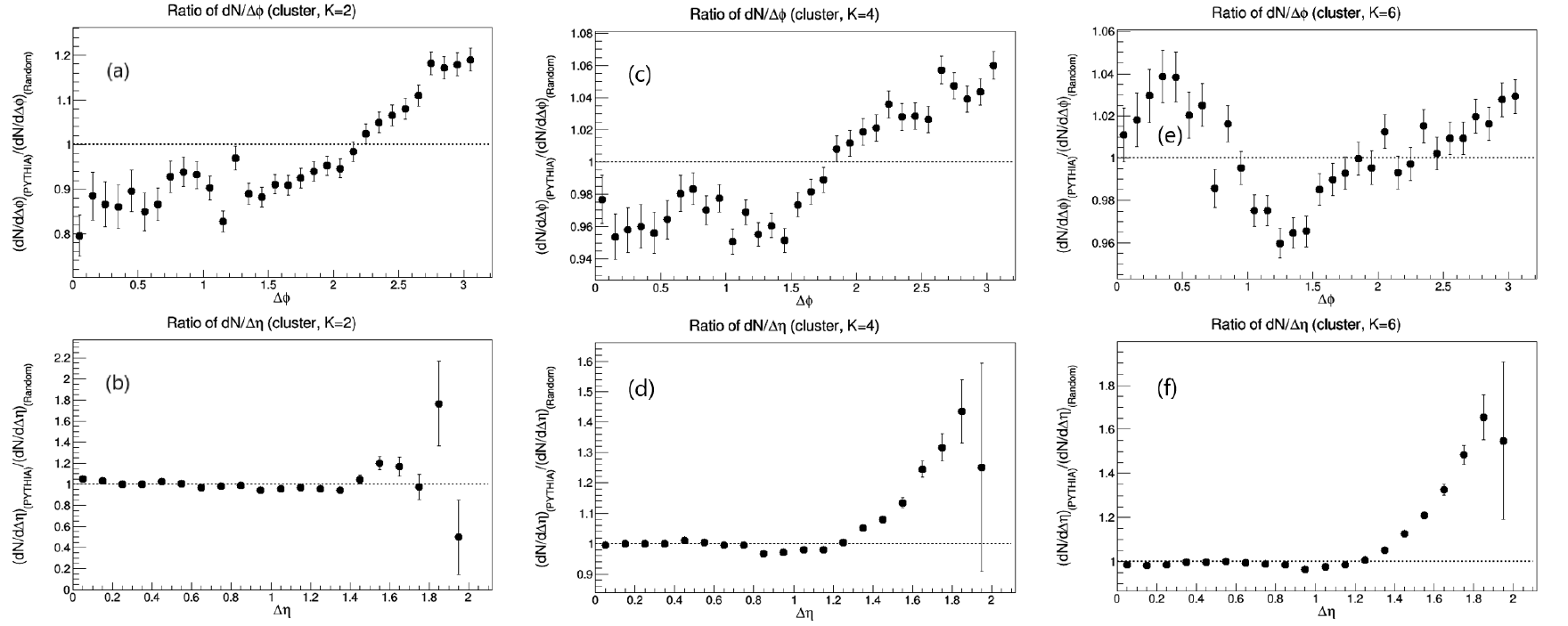}
\caption{Normalized two-clusters correlation distribution. The figures are produced by PYTHIA  8.1 generated and random events with 2, 4 and 6 clusters. The Figs.(a), (c) and(e) are the ratio of $\Delta \phi$ with function $(dN/d\Delta \phi)_{(PYTHIA)}/(dN/d\Delta \phi)_{(Random)}$, and the Figs.(b), (d), and (f) are the ratio of $\Delta \eta$ with function $(dN/d\Delta \eta)_{(PYTHIA)}/(dN/d\Delta \eta)_{(Random)}$.}
\label{fig24n}
\end{figure*}

We show   the normalized $dN/d\Delta \phi$  
and $dN/d\Delta \phi$ for $K=2$
in Fig.\ref{fig24n}(a)  and Fig.\ref{fig24n}(b)  
for $K=4$ in Fig.\ref{fig24n}(c) and Fig.\ref{fig24n}(d), and for $K$=6 in   
Fig.\ref{fig24n}(e)  and Fig.\ref{fig24n}(f).  
While these distributions collaborate what one can find out about the shape of the distributions in the full two-dimensional 
$dN/d\Delta\eta d\Delta \phi$
distribution, it is difficult to 
 extract information on the two-dimensional   
distribution from the marginal distributions.  What can be stated is that as a function of increasing $K$ values, the normalized $dN/d\Delta\phi$ distribution has a peak at $\Delta \phi \sim \pi$  that 
remains for $K=4$ and $6$.



\end{document}